\documentclass[11pt, twocolumn]{article}
\usepackage[a4paper,
  top=2cm,bottom=2cm,
  left=1.5cm,right=1.5cm,
  columnsep=0.5cm]{geometry}

\usepackage[english]{babel}
\usepackage[utf8]{inputenc}
\usepackage{bm}
\usepackage{nicefrac,xfrac}
\usepackage{amsmath}
\usepackage{mathtools}
\usepackage[linesnumbered,ruled,vlined]{algorithm2e}
\usepackage[colorlinks=true,linkcolor=blue, allcolors=blue,hyperfootnotes=false]{hyperref}
\usepackage{xcolor}
\usepackage{braket}
\usepackage{amsthm}
\usepackage{amssymb}
\usepackage{textcomp, gensymb}
\usepackage{siunitx}
\usepackage[version=4]{mhchem}
\usepackage{url}
\usepackage{hyperref}
\usepackage{float}
\usepackage{placeins} 
\DeclareSIUnit \Ha {Ha}

\usepackage{booktabs}
\usepackage{soul}  
\usepackage{csquotes}
\usepackage{orcidlink}

\usepackage{subcaption}
\usepackage{titling}

\usepackage{cleveref}

\DeclareMathOperator{\arctantwo}{arctan2}

\usepackage{cuted} 
\usepackage[style=numeric-comp, sorting=none, eprint=false]{biblatex}

\renewbibmacro*{issue+date}{%
    \iffieldundef{issue}{}{%
        \printtext[parens]{\printfield{issue}}%
        \setunit*{\addspace}%
    }%
    \printtext[parens]{\printfield{year}} 
}

\renewbibmacro*{url+urldate}{}

\SetKwInput{KwHardware}{Hardware Resources}

\DeclareSIUnit\angstrom{\text {Å}}

\usepackage{setspace}
\usepackage{xcolor} 
\usepackage{soul}

\usepackage[at]{easylist}
\usepackage{versions}
\usepackage{authblk}

\makeatletter
\def\@fnsymbol#1{\ensuremath{\ifcase#1\or \dagger\or \ddagger\or
   \mathsection\or \mathparagraph\or \|\or **\or \dagger\dagger
   \or \ddagger\ddagger \else\@ctrerr\fi}}
    \makeatother

\newif\ifPrintOutline
\PrintOutlinefalse
\excludeversion{commentdone}
\ifPrintOutline
\includeversion{comment}
\else
\excludeversion{comment}
\fi

\title{Efficient Operator Selection and Warm‑Start Strategy for Excitations in Variational Quantum Eigensolvers}

\author[1*\textdagger]{Max Haas~\orcidlink{0009-0004-9414-5442}}
\author[1\textdagger]{Thierry N. Kaldenbach~\orcidlink{0009-0008-5607-4427}}
\author[2]{Thomas Hammerschmidt}
\author[1]{Daniel Barragan-Yani}

\date{}

\affil[1]{German Aerospace Center (DLR), Institute for Frontier Materials on Earth and in Space, Cologne, Germany}
\affil[2]{ICAMS, Ruhr University Bochum, Bochum, Germany}

\AtEveryBibitem{%
  \hypertarget{\thefield{entrykey}}{}%
}

\newenvironment{customabstract}{
    \begin{strip}
    \centering
    \begin{minipage}{0.85\textwidth} 
    \vspace{-1em}
}{
    \vspace{2em}
    \end{minipage}
    \end{strip}
}

\setstcolor{red}
\begin{document}

    \maketitle

    \begin{customabstract}
        We present a novel approach for efficient preparation of electronic ground states, leveraging the optimizer ExcitationSolve 
        \hyperlink{Jaeger2025Fast}{[Jäger et al.,~\textit{Comm.~Phys.}~(2025)]} and established variational quantum eigensolver-based operator selection methods, such as Energy Sorting (ES). By combining these tools, we demonstrate a computationally efficient protocol that enables the construction of an approximate ground state from a unitary coupled cluster ansatz via a single sweep over the operator pool. Utilizing efficient classical pre-processing to select the majority of relevant operators, this approach reduces the computational complexity associated with traditional variational quantum eigensolver (VQE) optimization methods. We further show that second-order Epstein–Nesbet (EN2) perturbation theory emerges as the first‑order Taylor expansion of our protocol in terms of a correlation measure, clarifying why our approach provides a more robust initial guess for the ground state in strongly correlated regimes. We also find that second-order M\o ller-Plesset perturbation (MP2) theory, which is widely used for unitary coupled cluster (UCC) initialization, performs worse than both EN2 and our protocol. Furthermore, we show that our method can be seamlessly integrated with one-variational-parameter couple exchange operators, thereby further reducing the number of required CNOT operations. Overall, we empirically observe a quadratic convergence speedup beyond state-of-the-art methods, advancing the preparation of high-fidelity electronic ground states -- one of the cornerstones of meaningful electronic structure calculations in the noisy intermediate-scale quantum computing (NISQ) era, and a prerequisite for fault-tolerant quantum computing (FTQC) algorithms such as quantum phase estimation.
        \\ \\
        ${}^{\text{*}}$Corresponding author email: \url{max.haas@dlr.de} \hfill 
        DOI: \href{https://doi.org/10.48550/arXiv.2602.10776}{10.48550/arXiv.2602.10776} \\
        ${}^{\text{\textdagger}}$These authors contributed equally.
    \end{customabstract}

    
    \onehalfspacing

        \section{Introduction}

The variational quantum eigensolver (VQE) \cite{Peruzzo2014_VQE} has emerged as a paradigmatic approach to solve the ground-state problem in quantum many-body systems. This hybrid algorithm, which combines the computational power of both quantum and classical computers, has been subject to intense scrutiny and debate over the past decade \cite{Cerezo2021_VQAs, Endo2021_HybridQC_Algorithms, Gonthier2022_MeasurementsAsRoadblock}. Despite its promise, VQE is hampered by several fundamental challenges that hinder its ability to achieve quantum advantage \cite{Cerezo2025_DoesProvableAbsence}. 

One of the primary obstacles to overcoming this limitation is the phenomenon of barren plateaus, where the algorithm fails to converge due to exponentially vanishing gradients of the loss function \cite{Cunningham2025_Investigating_BPs, liu2024mitigating, Wang2021_noiseinducedbarren, Cerezo2025_DoesProvableAbsence, Uvarov_2021_OnBPs_in_VQAs}. This can be attributed to the exponential growth of the Hilbert space and the so-called curse of dimensionality. Furthermore, traditional ansätze such as the Unitary Coupled Cluster Singles Doubles (UCCSD) \cite{xia2020qubit} ansatz suffer from polynomial circuit depth scaling with respect to the number of electrons, resulting in deep circuits that are difficult to implement on current noisy intermediate-scale quantum (NISQ) devices. 

To address these challenges, researchers have been exploring novel strategies, including different ansatz designs \cite{Anastasiou2024Tetris-ADAPT, liu2024mitigating, park2024hardwareefficientansatzbarrenplateaus}, optimization techniques  \cite{nakanishi_SequentialMinimalOptimization_2020, Ostaszewski2021structure, Kottmann2021Feasible, Jaeger2025Fast} and initialization strategies \cite{Puig2025warmStarts, truger2024warmstartingvqeapproximatecomplex, Romero2018Strategies}. One promising approach is the use of adaptive ansätze such as ADAPT-VQE \cite{Grimsley2019_ADAPT-VQE,tang2021QubitADAPT, Grimsley2023adaptMitigatesBPs}, which employs an iterative process to add operators one by one from a pool of possible choices. However, this approach comes at the cost of high computational overhead, caused by the evaluation of an excessive number of quantum circuits during the selection process. 

Energy Sorting (ES) \cite{2023_Fan_energy_sorting} offers a solution to mitigate this limitation by selecting multiple operators simultaneously. By ranking all operators based on their impact on the energy and appending them to the ansatz in order, ES reduces the computational overhead associated with sequential operator addition. Moreover, ES can be parallelized to multiple quantum processing units (QPUs), thereby offering potential speedup opportunities. 

In this work, we build upon the idea of ES by combining it with the recently developed optimizer ExcitationSolve \cite{Jaeger2025Fast}. Since ExcitationSolve reconstructs the cost function for each parameter, it enables the direct optimization of parameters without gradient descent. This allows us to perform energy sorting \enquote{for free} in the first selection step. 

We demonstrate the efficacy of this approach using UCCSD ansätze for a range of molecules, from 4 to 20 qubits and compare its performance with fixed and adaptive ansätze in terms of operator count and evaluation requirements on the quantum computer. Our results show that we can select all relevant operators within a single operator selection step. As already mentioned in the original publication \cite{2023_Fan_energy_sorting}, the first selection process is classically simulable, leaving us with a fully classical method to construct and warm-start the VQE. 

We further explore the connection between our approach and more widely employed initialization strategies based on perturbative methods such as second order M{\o}ller-Plesset (MP2) \cite{MP1934, Cremer2011, Romero2018Strategies} or Epstein-Nesbet (EN2) perturbation theory \cite{Epstein1926, Nesbet1955}. In particular, we theoretically motivate and numerically confirm that ES provides the best initialization across the three methods in strongly correlated systems.

Notably, this approach can be adapted to accommodate the recently introduced One Variational Parameter Couple Exchange Operators (OVP-CEOs) \cite{ramoa2025reducing}. By adopting an OVP-CEO pool, we can further reduce the circuit depth compared to traditional ADAPT-VQE, at the cost of a small computational overhead.
    \section{Results}

    \subsection{Energy sorting with ExcitationSolve}\label{sec:ES}
    
        The ES algorithm works by simultaneously selecting all operators $U_j$ whose energy impact, $\Delta E_j=E_{\text{ref}}-E_j$, exceeds a predetermined threshold value $\epsilon_A$. $E_{\text{ref}}$ is the reference energy of the system before appending any operator and $E_j$ is the energy after appending $U_j$ to the ansatz and optimizing the corresponding parameter $\theta_j$. The optimizer ExcitationSolve \cite{Jaeger2025Fast} provide efficient means of calculating these energy differences. By fully reconstructing the energy landscape of each operator, ExcitationSolve can identify all operators fulfilling the condition with a single sweep over the operator pool, without requiring further optimization. Consequently, when employing ExcitationSolve, ES can be performed with no additional computational effort compared to a single adaptive step.
        When the threshold $\epsilon_A$ is set to 0, all operators contributing to the system’s energy are selected simultaneously. These operators can then be sorted by their impact and appended to the circuit. Moreover, ExcitationSolve also provides the optimal parameter value for each operator, enabling a warm start by initializing the parameters $\theta_i$ to finite values rather than the commonly chosen $\theta_i=0$.  The resulting ansatz can then be optimized like a fixed ansatz VQE, eliminating the need for subsequent operator selection. For linear systems like the \ce{LiH} molecule, the ansatz only needs to be constructed once and can be reused for any bond lengths, because the dissociation length does not change which orbitals overlap and therefore the relevant operators remain the same. Examples are given in Supplementary Material C.

    \subsection{Energy sorting for OVP-CEOs}\label{subsec:theo_ES_and_CEOs}
         
        The ExcitationSolve algorithm is applicable to any pool, $O=\{G_1, G_2, \dots, G_M\}$ comprised of generators $G$ that satisfy the condition $G^3=G$. A recently developed class of operators fulfilling this condition are one-variational-parameter couple exchange operators (OVP-CEOs) \cite{ramoa2025reducing}; a formal proof of this relationship is provided in Supplementary Material A. The OVP-CEOs generators are given by
        \begin{equation}
        \begin{split}
            &G_{\alpha_1\beta_1\alpha_2\beta_2}^{(\text{OVP-CEO},+)} \coloneqq G_{\alpha_1\beta_1\rightarrow\alpha_2\beta_2}^{(\text{QE})}+G_{\alpha_2\beta_1\rightarrow\alpha_1\beta_2}^{(\text{QE})} \\
            &=Q_{\alpha_2}^\dagger Q_{\beta_2}^\dagger Q_{\alpha_1}Q_{\beta_1}+Q_{\alpha_1}^\dagger Q_{\beta_2}^\dagger Q_{\alpha_2}Q_{\beta_1}- \textrm{H.c.},
        \end{split}
        \end{equation}
        and
        \begin{equation}
        \begin{split}
            &G_{\alpha_1\beta_1\alpha_2\beta_2}^{(\text{OVP-CEO},-)}\coloneqq G_{\alpha_1\beta_1\rightarrow\alpha_2\beta_2}^{(\text{QE})}-G_{\alpha_2\beta_1\rightarrow\alpha_1\beta_2}^{(\text{QE})}  \\
            &=Q_{\alpha_2}^\dagger Q_{\beta_2}^\dagger Q_{\alpha_1}Q_{\beta_1}-Q_{\alpha_1}^\dagger Q_{\beta_2}^\dagger Q_{\alpha_2}Q_{\beta_1}-\textrm{H.c.},
        \end{split}
        \end{equation}
        with the qubit creation- and annihilation operators
        \begin{align}
            Q^\dagger_i=\frac{1}{2}\left(X_i-iY_i\right), \qquad  Q_i=\frac{1}{2}\left(X_i+iY_i\right).
        \end{align}
        Key to their functionality are the terms $Q_{\alpha_1}^\dagger Q_{\beta_2}^\dagger Q_{\alpha_2}Q_{\beta_1}$ and their Hermitian conjugates (H.c.), which mix the excitation on the $\alpha$-orbital with the de-excitation on the $\beta$-orbital, and vice versa. This mixing reduces the quantum cost of implementing each excitation operator from a count of 13 CNOT gates for standard qubit excitations \cite{Yordanov2020_EfficientQuantumCircuits, Yordanov2021_QubitBasedVQE} to 9 CNOT gates with OVP-CEOs, and a depth of 11 to 7.\\
        
        However, the application of ExcitationSolve to a pool of OVP-CEO$^+$ and OVP-CEO$^-$ operators presents two challenges. First, the resulting variational ansatz is twice as deep as necessary. This arises because both OVP-CEO$^+$ and OVP-CEO$^-$ act like a double excitation when applied to the HF state, leading to redundant selection of either both or neither operator. This negates the gate count reduction afforded by OVP-CEOs, resulting in circuits even deeper than those employing a UCCSD pool.  Restricting the pool to solely OVP-CEO$^+$ or OVP-CEO$^-$ operators is insufficient to reach the same convergence, as demonstrated in \Cref{subsec:exp_ES_and_CEOs}. 
        Second, a naive selection of OVP-CEOs based on their impact on the HF state leads to an increased initial energy. This is attributed to the orbital mixing terms within the OVP-CEO formulation, which are inactive on the HF state due to the virtual nature of the orbitals involved. However, subsequent application of other OVP-CEOs that occupy these virtual orbitals introduces a non-zero contribution from the de-excitation terms. To address this, an additional selection step is implemented that, for each pair of OVP-CEO$^+$ or OVP-CEO$^-$ acting on the same orbitals, determines which operator will be appended to the ansatz. The algorithm initiates with the pure OVP-CEO$^+$ pool, and all relevant operators are selected using ExcitationSolve, but are not immediately appended to the ansatz. Instead, they are ordered according to their predicted impact on the Hartree-Fock (HF) state. Subsequently, for each operator a determination is made as to whether the OVP-CEO$^+$ or OVP-CEO$^-$ variant is more advantageous and the selected operator is then added to the ansatz. This process is repeated for each operator in the ordered list. For the initial operator selection, the choice is arbitrary and a distinction only exists due to numerical inaccuracies, as both OVP-CEO$^+$ and OVP-CEO$^-$ exhibit the same energy impact on the HF state. However, for operators added subsequently, the state upon which they act is no longer the HF state, resulting in potentially significant differences in their energy contributions. In each case, the operator with the greater predicted impact was chosen.
    
    \subsection{Classical simulation of the operator selection via energy sorting} \label{sec:Classical_simulation}
    
        When building the ansatz using ES from either a UCC or OVP-CEO pool, a large part of the operator selection can be performed classically, saving valuable quantum resources and giving a warm-start to the quantum simulation.
        
        When applying a UCC ansatz on the reference state obtained from Hartree-Fock theory, one typically starts by considering all spin-preserving double excitations from occupied to virtual orbitals. 
        
        The contributions of higher-order excitations, i.e., triples or higher, can always be ruled out as the electronic structure Hamiltonian only entails quadratic and quartic fermionic terms and therefore cannot couple two classical states that differ by more than a double-excitation \cite{Shkolnikov2023avoidingsymmetry}.
        It is proven that single excitations cannot lower the energy when applied to the HF ground state as it is the classical state with the energy closest to the true ground state. Thus, having a single-excitation lowering the energy would imply the existence of a different classical state with lower energy than the HF state \cite{Shkolnikov2023avoidingsymmetry}.
        
        The selection of the double excitations (or OVP-CEOs) with ES works by testing each operator individually for its energy impact on the HF state, as described in detail in \Cref{sec:ES}. The states for which the energy must be determined are thus a classical state with exactly one double excitation on top. Such unitary double excitation then spans the subspace of the HF state and the double-excited state, an effective two-state system.  
        It is convenient to interpret the action of the double excitation on the reference state in the qubit picture instead of the fermionic picture. While a double excitation typically maps to eight Pauli rotations under the Jordan-Wigner mapping, the circuit can be reduced down to a single Pauli rotation of the type $\exp(\pm i\theta X_p X_q X_r Y_s)$ by exploiting the structure of the initial state (cf.~Supplementary Material \ref{sec:proof_double_excitation_to_pauli}). This can serve as a helpful circuit optimization technique to assemble the beginning of the circuit, as it has been demonstrated in Ref.~\cite{Jaeger2025Fast}. One could then map this circuit to a measurement pattern entailing only one qubit \cite{Kaldenbach2024Mapping}, which aligns with the fact that we are dealing with a two-state system.
        
        It turns out that there is no need to simulate even this single-qubit system. Instead, 
        by reducing the electronic structure Hamiltonian to that subspace, we find an analytical form to calculate the largest energy reduction
        \begin{align}
            \Delta E_\textrm{min} = a - \sqrt{a^2+b^2} \leq 0,
            \label{eq:energy_impact_HF}
        \end{align} 
        and corresponding optimal initial parameters, which we refer to as ES guess amplitudes,
        \begin{align}
            \theta_\textrm{ES} = -\frac{1}{2}\arctan\left(\frac{b}{a}\right),
            \label{eq:optimal_angle_HF}
        \end{align}
        where the parameters $a$ and $b$ are directly obtained from one- and two-electron integrals for each double excitation in the operator pool. More specifically, the parameter $a$ is proportional to the energy difference between the excited state and the reference state. For Hartree-Fock calculations, the energy of any excited state will always be larger than the Hartree-Fock ground state energy, and hence we have $a>0$. Therefore, energy reductions from Eq.~\eqref{eq:energy_impact_HF} can be entirely attributed to the parameter $b$ which entails the coupling between the reference state and the double-excited state. 
        For the derivations of the energy impact and optimal angle, as well as the quantitative definitions of the parameters $a$ and $b$ in terms of one- and two-electron integrals, refer to Supplementary Material \ref{sec:proof_classical_simulation_doubles}.  

        Using these equations, one can classically compute the energy impact of each double excitation and the warm start parameters. 
        For a rigorous proof of the energy impact and circuit simplifications, refer to Supplementary Material B.
        
        After the first layer of double excitations, the state against which all other orders of excitations are tested is no longer a classical state. The presented formulas and circuits then no longer hold and the operator selection itself requires the quantum device.
        
        Last, it is worth highlighting that if the reference state is not obtained from a Hartree-Fock calculation, but for example from Kohn-Sham theory, as observed in UCC-based post-DFT calculations \cite{ma2020quantum, schultheis2025manybody}, single excitations can potentially lower the energy compared to the reference state. For that case, we also provide analytical formulas and circuit simplifications in Supplementary Material \ref{sec:proof_classical_simulation_singles}. Then the selection and initialization of single excitations can be done in the same fashion as described for doubles. Note that within this work, we only perform post Hartree-Fock calculations and do not employ this technique.   
        
    \subsection{Comparison to operator selection based on perturbation theories}
    
    It has long been recognized that for UCC simulations designed to achieve rapid convergence, it is best not to initialize them naively with 0, but instead with physically motivated amplitudes \cite{Romero2018Strategies}. Second-order perturbation theories are used for this purpose, whereby the choice of the unperturbed and perturbed Hamiltonians is initially arbitrary. From the terms contributing to the second-order energy correction, so-called guess amplitudes \cite{Romero2018Strategies} can be derived, which are then used as initialization parameters for the excitation. For more details on second order perturbation theory in the context of electronic structure, and the extraction of guess amplitudes from the perturbative energy corrections, refer to Supplementary Material \ref{sec:pert_theories}. 

    In the UCC context, the Møller–Plesset second-order (MP2) perturbation theory \cite{MP1934, Cremer2011} is particularly widespread. Here, the effective single-particle Hamiltonian is regarded as the unperturbed Hamiltonian, whilst two-particle terms are treated as perturbations. From this, the MP2 guess amplitudes for double excitations can be derived. Unlike the ES amplitudes, the MP2 amplitudes depend on the differences between the occupied orbitals energies before and after the excitation, but not on the energy difference between the reference state and the excited state. Consequently, the changes in Coulomb repulsion energies when exciting electrons are not taken into account in the amplitude, which leads to a poorer initial guess even in weakly correlated regimes, as the full physics is not incorporated. For a detailed recap of the MP2 amplitude equations and a quantitative comparison to the ES amplitudes, consider Supplementary Material \ref{subsec:MøllerPlesset}.

\begin{figure*}[t]
        \centering
        \includegraphics[width=\textwidth]{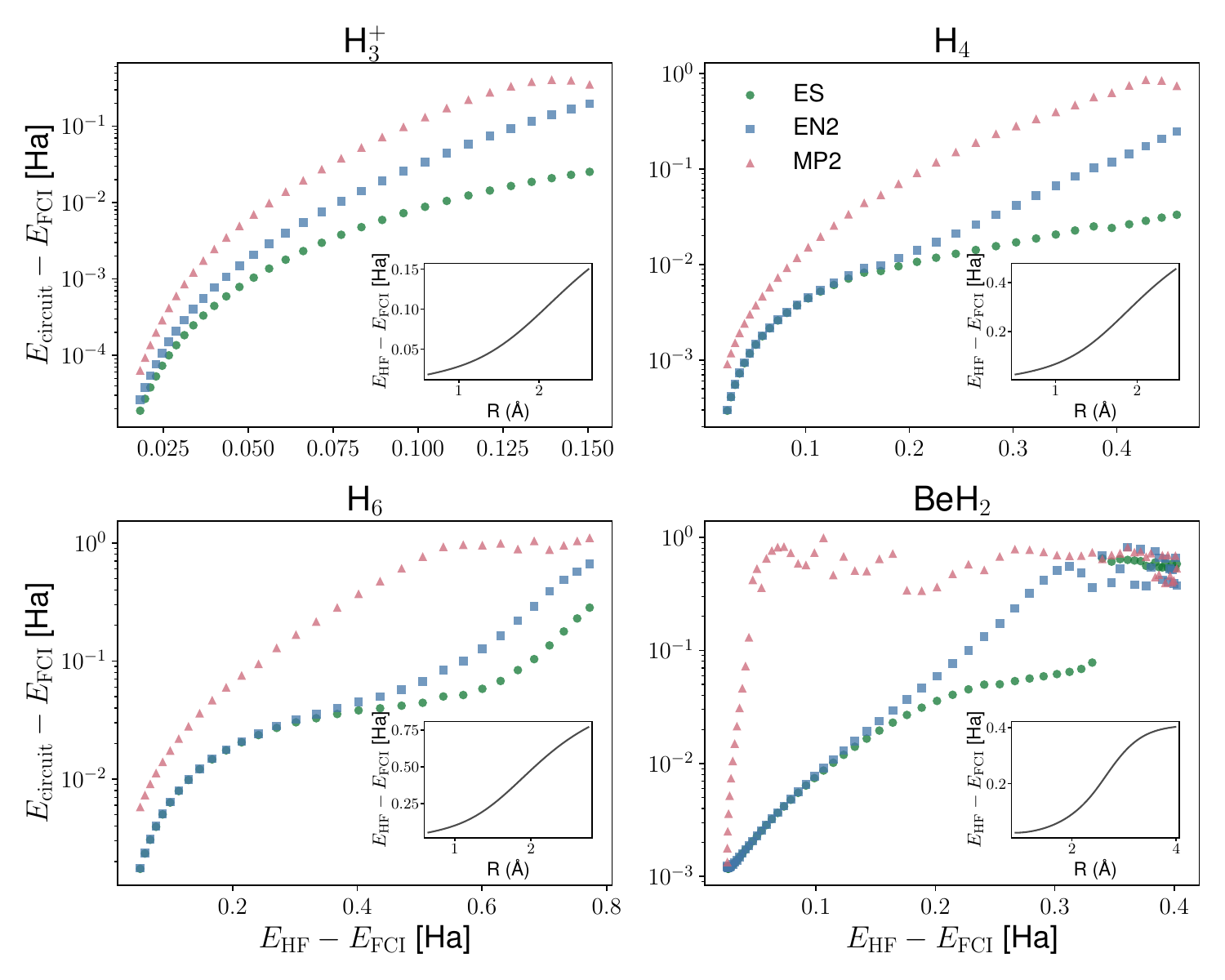}
        \caption{Error to FCI when initializing a UCCD circuit with parameters predicted using MP2, EN2 and ES over electron-electron correlation. The strength of the correlation is measured by how close the HF energy approximates the FCI and is dependent on the bond distance (correlation over bond distance in the insets). Stronger correlation leads to larger exchange terms, which are not accounted for by MP2 and approximated in EN2, resulting to a better performance of ES in this regime.}
        \label{fig:GuessAmplitudesComparison}
    \end{figure*}

    Less common in the UCC context, but much more closely related to our ES method, is the Epstein–Nesbet second-order (EN2) perturbation theory \cite{Epstein1926, Nesbet1955}. Here, the unperturbed Hamiltonian comprises only the diagonal terms, whilst all single and double excitations are treated as perturbations. This means that the EN2 guess amplitudes, just like ES, take the Coulomb repulsion into account. In Supplementary Material \ref{subsec:EpsteinNesbet} we show that the EN2 amplitudes can be derived directly from the ES amplitudes as a first-order Taylor series expansion in the correlation measure $b/a$:
    \begin{align}
        \theta_\textrm{ES}
        = \theta_\textrm{EN2} + \mathcal{O}((b/a)^2).
        \label{eq:ES_EN2_Amplitudes_Taylor}
    \end{align}
    Consequently, both initialization methods are almost identical in the region of weak correlation. In Figure.~\ref{fig:GuessAmplitudesComparison}, we compare all three initialization methods (ES, MP2, EN2) by initializing the parameters of a UCCD ansatz with the corresponding guess amplitudes. We then compute the energy error to the FCI energy to assess the quality of the initialization. To study the impact of the electronic correlation, we stretch various molecules which increases the correlation energy. We find that MP2 consistently provides the worst initial guess across all geometries/correlations. As predicted by Eq.~\eqref{eq:ES_EN2_Amplitudes_Taylor}, the ES and EN2 amplitudes lead to almost identical circuit parameter initialization and energies in low correlation regimes, but then drift apart for strong correlations at stretched geometries. Importantly, we find that ES stays robust across a wider range of geometries as the correlation increases, and provides the best initial guess overall. Finally in the example of \ce{BeH2} all three methods fail as the molecule dissociates. This is however not a limitation of ES, but rather the self-consistent HF calculation which fails as molecular orbitals are no longer a fitting description of the system. Overall, while not exhibiting a large improvement over small molecules in equilibrium geometry, we observe that using ES has an edge over the perturbation theories across the studied strongly correlated systems.
    
    
    When simulating molecules using a UCCSD ansatz, an inverse relation exists between the size of the simulated system and the precision of the converged energy obtained with all optimizers employed. For \ce{H2} (4 qubits) most optimizers achieve an energy value that is closer than $\SI{1e-13}{\Ha}$ from the FCI solution, while for \ce{LiH} (12 qubits), convergence plateaus at $\SI{1e-5}{\Ha}$ and for \ce{H2O} (14 qubits) at $\SI{1e-4}{\Ha}$. Therefore, to achieve a minimum precision of $\SI{1e-3}{\Ha}$ to the FCI solution (often referred to as chemical accuracy), for larger molecules the ansatz must be expanded to include higher order excitations to improve convergence \cite{Haidar2023_UCCSDT}. Fortunately all fermionic or qubit excitation operators of any order satisfy the necessary condition of $G^3=G$, rendering them compatible with ExcitationSolve \cite{Jaeger2025Fast}. However, analogous to single excitations, they cannot improve upon the HF state \cite{Shkolnikov2023avoidingsymmetry} and must therefore be selected after the double excitations have been appended to the ansatz. Consequently, their selection can not be simulated efficiently classically as the reference state is no longer a classical state.
    \Cref{exp:higher_order_excitations} explores the convergence behavior when utilizing triple excitations for the example molecule \ce{LiH}.
    
\subsection{Energy sorting using the UCCSD pool}

    Starting from a UCCSD pool of operators, an ansatz is constructed as explained in \Cref{sec:ES}. The initial state is chosen to be the HF state and $\epsilon_{\textrm{A}}$ was set to $10^{-13}\,\si{\Ha}$ to avoid the inclusion of contributions arising from numerical imprecision. Given that single excitations cannot lower the energy in a HF state \cite{Shkolnikov2023avoidingsymmetry}, a sequential approach was adopted; relevant double excitations were initially identified and appended to the ansatz, followed by the selection and inclusion of single excitations. The selection of the double excitations can efficiently be performed on a classical computer as explained in \Cref{sec:Classical_simulation}, while the selection of the single excitations and the following optimization have to be performed on a quantum device. In this case, the quantum device is simulated by a noiseless state-vector simulator provided by the TenCirChem package \cite{Li2023_TenCirChem}. Finding the optimal parameter value for a single operator $U_j$ using ExcitationSolve requires the evaluation of the cost function given by \cref{eq:costfunction} at four different values of $\theta_j$, which will be referred to as energy evaluations in the following text and figures.

    The efficacy of this algorithm, coupled with an Energy Sorting strategy (ExcitationSolve + ES), was evaluated for the molecules \ce{H2}, \ce{H3^+}, \ce{He2}, \ce{OH-}, \ce{LiH}, \ce{H6}, \ce{H2O}, \ce{BeH2}, \ce{NH3}, \ce{BH3}, \ce{CH4} and \ce{C2}, each modeled in their respective equilibrium geometry as given in the datasets \cite{Utkarsh2023Chemistry_H2, Utkarsh2023Chemistry_H3plus, Utkarsh2023Chemistry_He2, Utkarsh2023Chemistry_OHminus, Utkarsh2023Chemistry_LiH, Utkarsh2023Chemistry_H6, Utkarsh2023Chemistry_H2O, Utkarsh2023Chemistry_BeH2, Utkarsh2023Chemistry_NH3, Utkarsh2023Chemistry_BH3, Utkarsh2023Chemistry_CH4, Utkarsh2023Chemistry_C2}.  The convergence behavior of ExcitationSolve + ES was benchmarked against the original ADAPT-VQE algorithm (GD + gradient selection), and adaptive ExcitationSolve without the use of ES (Exc.Solve + Exc.Solve selection). For \ce{NH3}, \ce{BH3}, \ce{CH4} and \ce{C2} it was not feasible to simulate ADAPT-VQE due to its high computational runtime, for \ce{C2} even Exc.Solve + Exc.Solve selection was too costly.
    
\begin{figure*}[p]
        \centering
        \includegraphics[width=\textwidth]{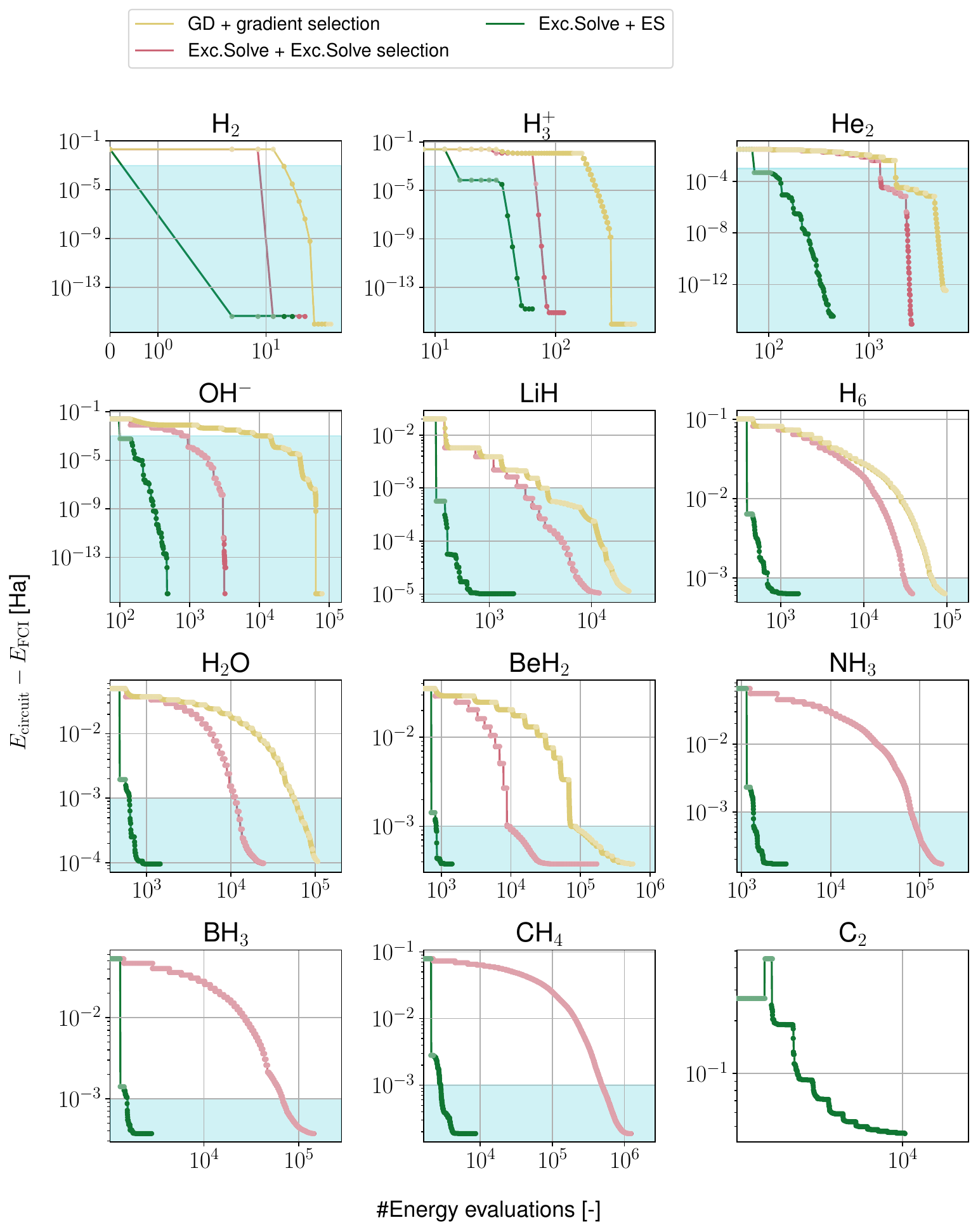}
        \caption{Comparison of ExcitationSolve + ES (green) to ADAPT VQE (yellow) and naive adaptive ExcitationSolve (red). Lighter colors signal quantum resources spent on operator selection, darker colors mean quantum resources spent on VQE optimization. The light blue area marks chemical accuracy.}
        \label{fig:ES_vs_ADAPT}
    \end{figure*}

    In \Cref{fig:ES_vs_ADAPT} the convergence profiles of the different studied algorithms are presented. All curves utilize a dual-color scheme. Lighter shades indicate computational resources dedicated to operator selection, while darker shades represent resources allocated to variational quantum eigensolver (VQE) optimization. The results demonstrate that naive Exc.Solve + Exc.Solve selection is dominated by the operator selection process. In contrast, the ES strategy enables operator selection in a single sweep, followed by a limited number of VQE optimization iterations. Following this initial construction, a final screening process verifies the absence of additional operators exceeding the $E_{\textrm{th}}$ threshold.  Consequently, the combination of ES and ExcitationSolve facilitates the construction of a compact ansatz comprised solely of operators contributing significantly to the ground state, and provides a warm-start to the optimization process by initializing each operator with its optimal parameter $\theta_j$ relative to the initial state.

    \begin{figure*}[t]
        \centering
        \includegraphics[width=1\textwidth]{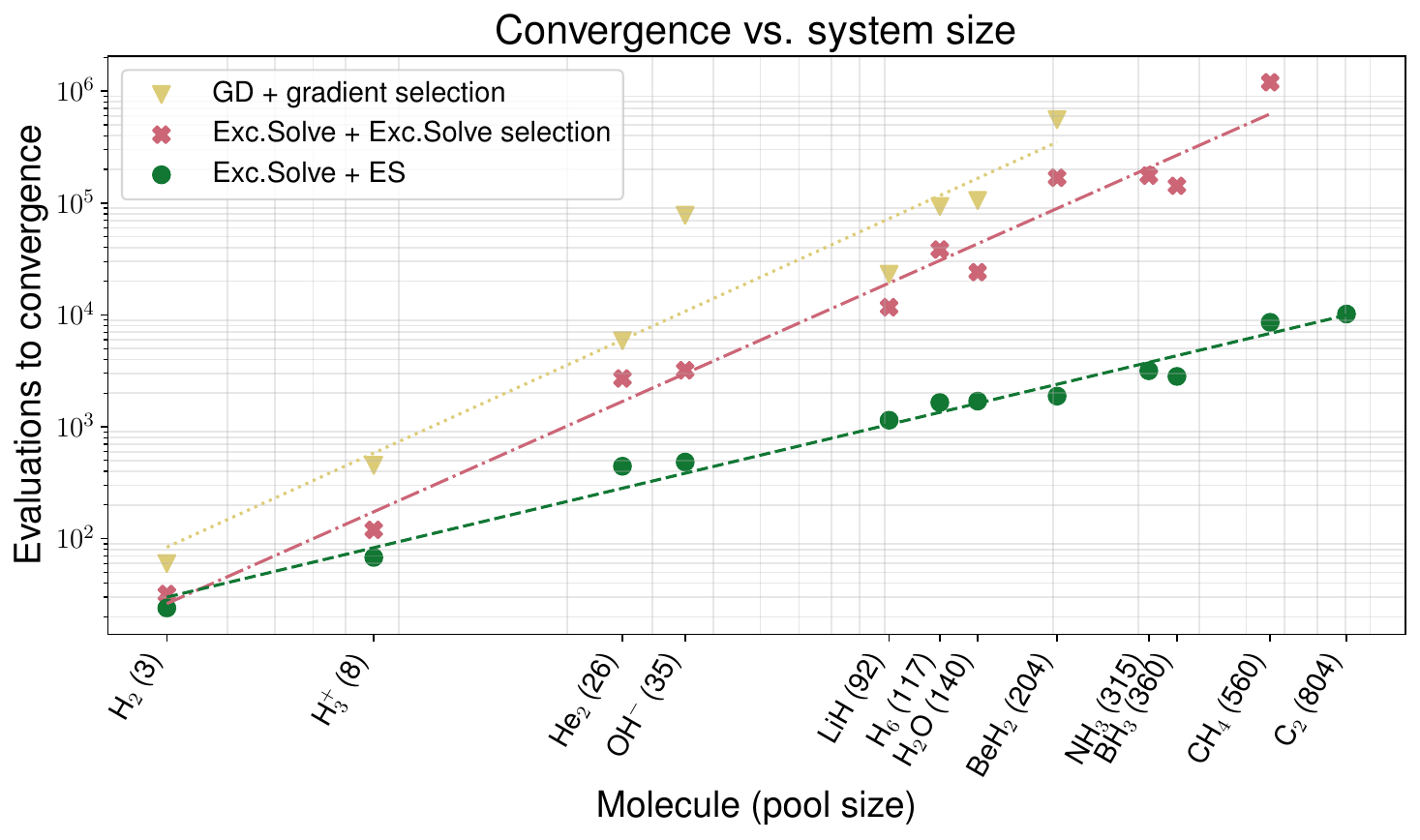}
        \caption{Evaluations to convergence over pool size on log-log scale, comparing ExcitationSolve + ES (green) to ADAPT VQE (yellow) and adaptive ExcitationSolve (red). An unweighted linear fit to the log-log data is used as guide to the eye. As the pool size grows, the number of evaluations required on the QC grows exponentially in all cases, but the exponent when using ExcitationSolve + ES is reduced significantly.}
        \label{fig:speed-up}
    \end{figure*}
    
    Another trend is visible in \Cref{fig:ES_vs_ADAPT}: for \ce{LiH} with an operator pool of 92 operators ExcitationSolve + ES requires about $10^{-1}$ of the resources of pure ExcitationSolve, while for \ce{CH4} with a pool size of 804 the cost is reduced by a factor of $10^{-2}$. \Cref{fig:speed-up} formalizes this trend and makes it clearly visible. Here the number of evaluations required to reach convergence is plotted over the respective pool size for each molecule on a log-log scale. Indeed, the computational resources required to run the VQE grows exponentially with the number of operators in the UCC pool for all methods. While using Exc.Solve + Exc.Solve selection gives a constant speed-up over ADAPT-VQE using GD as suggested in \cite{Jaeger2025Fast}, using  ExcitationSolve + ES even reduces the exponent significantly. An unweighted linear fit to the log-log data is used as guide to the eye. The slope is reduced approximately by a factor of 2 for ExcitationSolve + ES compared to both GD + gradient selection and Exc.Solve + Exc.Solve selection, so the speed-up is quadratic in the pool size.
    
\begin{figure}
        \centering
        \includegraphics[width = 0.8\linewidth]{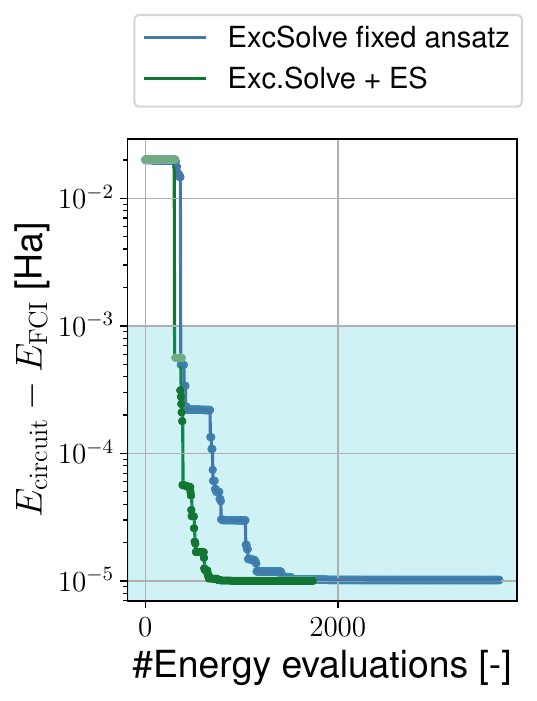}
        \caption{Fixed ansatz (blue) convergence compared to ExcitationSolve + ES (green) for the \ce{LiI} molecule. Even though the convergence process begins earlier for the fixed ansatz, the optimization using ES is faster, because all unnecessary operators are removed from the ansatz. The light blue area marks chemical accuracy.}
        \label{fig:ES_vs_fixed}
    \end{figure}

    Exc.Solve + ES can be viewed as a method to efficiently construct a fixed ansatz based on a UCC pool, which removes all irrelevant operators from the ansatz and warm starts the simulation. \Cref{fig:ES_vs_fixed} compares the convergence of the ExcSolve + ES algorithm to a fixed UCCSD ansatz for the \ce{LiH} molecule.  
    The Exc.Solve + ES approach incurs an initial computational cost associated with operator selection, while the fixed ansatz immediately starts its optimization. However, the substantial number of operators within the fixed ansatz (96 for \ce{LiH}), many of which exhibit minimal or no impact at all, significantly slows down the optimization process and leads to plateaus in the optimization curve. Once constructed, the ansatz generated by ES contains fewer operators (34 for \ce{LiH}), making it more shallow and converging rapidly, outperforming the fixed UCCSD ansatz.
    
\subsection{Higher order excitations}\label{exp:higher_order_excitations}

    As the number of electrons in the systems studied in \Cref{fig:ES_vs_ADAPT} increases, the convergence of the VQE is increasingly limited by the order of excitations permitted in the ansatz. While a precision of $10^{-15}\,\si{\Ha}$ can be achieved for \ce{H2}, calculations for \ce{H2O} are limited to $10^{-4}\,\si{\Ha}$, and for \ce{C2} the results no longer even reach chemical accuracy. Consequently, scaling up simulations requires not only accounting for the $\mathcal{O}(n^4)$ growth in the number of double excitations, with $n$ being the number of electrons in the system, but also incorporating higher-order excitations to attain the desired precision. Converging a VQE with a fixed ansatz entailing triple-, or higher order excitations quickly becomes computationally intractable due to the rapid increase in circuit depth with each added order of operator, i.e., the number of Pauli terms in the excitation generators grows exponentially with the order of the excitations.
    
\begin{figure}
        \centering
        \includegraphics[width = 0.8\linewidth]{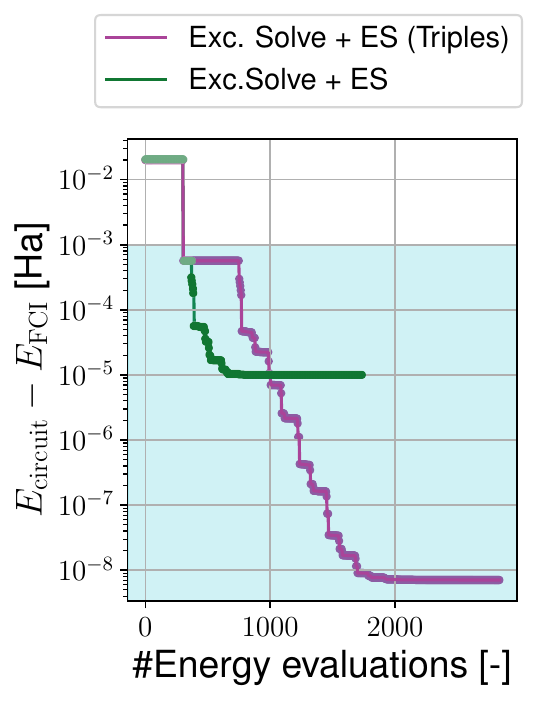}
        \caption{ExcitationSolve and Energy Sorting for a UCCSDT pool (purple) for a \ce{LiH} molecule compared to a UCCSD pool (green). First double excitations are appended to the ansatz, then singles and lastly triples, followed by the optimization of the complete circuit. The light blue area marks chemical accuracy.}
        \label{fig:triple_excitations}
    \end{figure}

    Fortunately, any order of fermionic or qubit operator can be treated with ExcitationSolve, enabling the ES protocol to include only the necessary operators in the ansatz. \Cref{fig:triple_excitations} demonstrates the improved convergence achieved with triple excitations for the \ce{LiH} molecule. The process starts again by selecting only double excitations, as these are the only operators that directly impact the HF state. Once the relevant double excitations have been appended to the ansatz, single and triple excitations are selected. This selection process is evident in the extended plateau observed around 500 evaluations in \Cref{fig:triple_excitations}. The resulting ansatz is then optimized using ExcitationSolve. This approach achieves a precision more than two orders of magnitude better than using only double excitations, at the cost of approximately three times the number of evaluations and the inclusion of 10 triple excitations. 

\subsection{Energy sorting using OVP-CEOs}\label{subsec:exp_ES_and_CEOs}
    
    The OVP-CEOs were implemented according to ref.~\cite{ramoa2025reducing}. An operator pool, $O_{\textrm{CEO}}$, was then constructed comprising all OVP-CEO$^+$ and OVP-CEO$^-$ operators, in addition to single excitations. Given that one OVP-CEO$^+$ and one OVP-CEO$^-$ operator exist for each double excitation within the UCCSD pool, the size of the $O_{\textrm{CEO}}$ pool is approximately double that of the UCCSD pool. Operator selection and ansatz optimization were performed using the ExcitationSolve algorithm. 

\begin{figure}
        \centering
        \includegraphics[width = 0.8\linewidth]{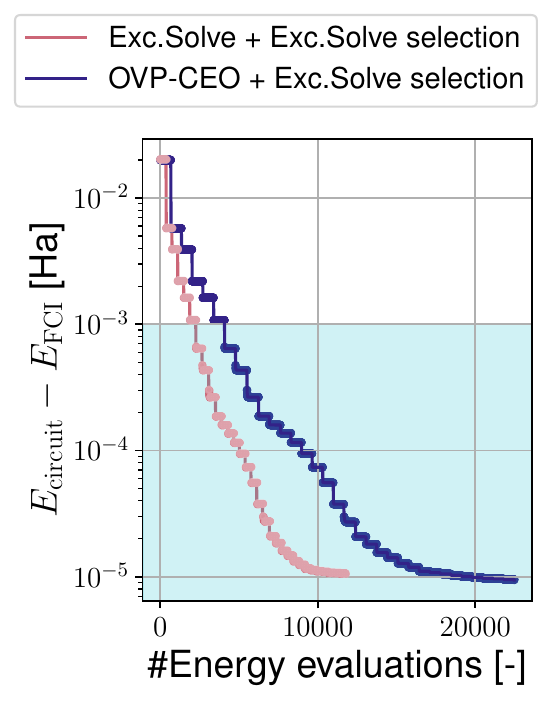}
        \caption{Adaptive optimization of an OVP-CEO pool (blue) compared to a UCCSD pool (red) of \ce{LiH}. Convergence is reached with the same number of operators, but for OVP-CEOs twice as many evaluations are needed due to the increased pool size. The light blue area marks chemical accuracy.}
        \label{fig:CEO_Adapt}
    \end{figure}

    \Cref{fig:CEO_Adapt} shows the convergence of an adaptive ansatz utilizing OVP-CEOs, compared to that employing the UCCSD pool \cite{Utkarsh2023Chemistry_LiH}, for the LiH molecule. The optimizer converges to the same ground state energy with the same number of operators. However, the larger pool size necessitates approximately twice the number of energy evaluations, increasing the computational cost of operator selection. This represents a trade-off between the depth of the resulting quantum circuit and the number of evaluations required. 
    
\begin{figure}[t]
        \centering
        \includegraphics[width = 0.8\linewidth]{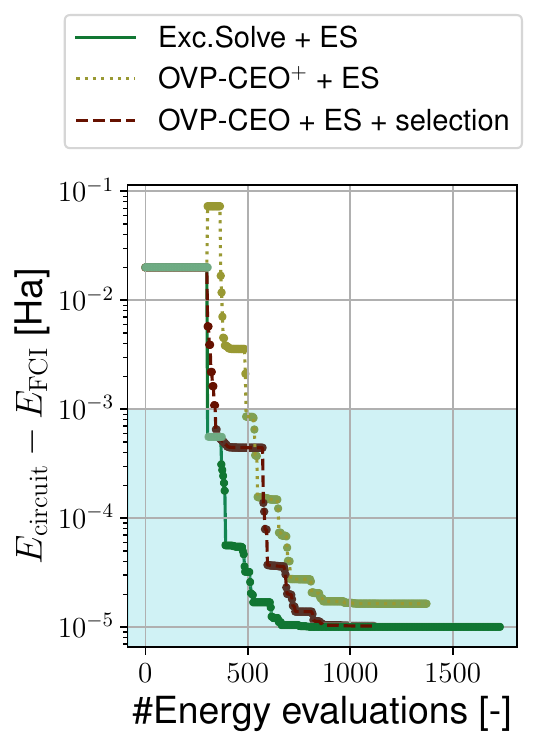}
        \caption{Combination of ES and OVP-CEOs. The light green curve shows the problems of a naive combination of a pool consisting of only OVP-CEO$^+$. With an additional selection step (brown) the performance can be improved to converge to the same energy as with the excitation operators (dark green). The dark green curve is the same as in \Cref{fig:CEO_ES} and serves as a reference. The light blue area marks chemical accuracy.}
        \label{fig:CEO_ES}
    \end{figure}

    ExcitationSolve was then applied to a pool consisting solely of OVP-CEO$^+$ operators and the convergence behavior was compared to that of a pool of qubit excitation operators, see \Cref{fig:CEO_ES}. The resulting ansatz comprised of OVP-CEO$^+$ operators exhibits insufficient expressivity to achieve the same level of precision upon convergence. To incorporate OVP-CEO$^-$ operators into the ansatz without doubling the circuit depth, an additional operator selection process was implemented as discussed in \Cref{subsec:theo_ES_and_CEOs}.
    
    This additional selection process increases the number of evaluations performed on the quantum computer, but enables the effective combination of ExcitationSolve with OVP-CEOs. Comparing the convergence behaviors presented in \Cref{fig:CEO_ES}, the use of OVP-CEOs with this selection criterion is found to be less than a factor of two slower than using excitation operators, while retaining the advantage of reduced circuit complexity – decreasing the number of CNOT gates per operator from 13 to 9, and depth from 11 to 7. For currently available NISQ hardware, where circuit depth is the limiting factor, this represents a favorable trade-off.

    \section{Discussion}

In this study, we employed the ExcitationSolve optimizer in conjunction with EnergySorting (ES) to develop an efficient hybrid method for constructing ansätze in quantum computing. Specifically, we combined the shallow circuits of ADAPT-VQE with the limited number of optimization steps from fixed ansatz optimization. By leveraging ES, which enables the computation of the change in energy caused by each operator within a single sweep over the operator pool, this approach eliminates the need for repeated evaluations of individual operators and significantly reduces the computational complexity associated with optimizing the ansatz. We even derived an analytical equation, which allows operator selection directly from Hamiltonian matrix elements. This approach enables the construction of an ansatz consisting solely of the most relevant operators, mirroring the strategy employed in ADAPT-VQE, but with reduced computational overhead. We demonstrated the efficacy of our method on twelve benchmark molecules, achieving a quadratic speed-up compared to the original ADAPT-VQE while maintaining equivalent circuit depth.  
We further refined our approach by adapting it to accommodate OVP-CEOs, a novel class of operators that can reduce circuit depth at the cost of increased operator pool size. This trade-off is particularly relevant in the context of noisy intermediate-scale quantum (NISQ) hardware, where circuit depth limitations are a significant challenge. Our method's performance was evaluated on the example of \ce{LiH}, yielding improved computational efficiency with minimal loss in precision.
The theoretical link between ES and perturbation theories, such as MP2 and EN2, shows that ES is not merely an efficient heuristic, but is also fundamentally intertwined with established perturbative methods. In particular, we were able to show that parameter initialization using EN2 theory effectively corresponds to a Taylor series expansion of ES in terms of electronic correlation. Both theoretically and numerically we show that ES is identical to EN2 in weakly correlated regimes, but represents a more robust and advantageous initialization in strongly correlated regimes.
The knowledge of all relevant operators before the ansatz construction leaves potential for further reduction of the circuit depth as algorithms like Tetris-ADAPT \cite{Anastasiou2024Tetris-ADAPT} could be employed to stack the operators in the most favorable way. 

As our method reduces compute times from days down to minutes (as in the example of \ce{CH4}), it provides an speed-up that makes out-of-the-box simulation of larger molecules feasible. While larger molecular simulations up to 28 qubits have already been demonstrated \cite{Cao2022_TowardLargerMolecularSimulation, Chawla2025_RelativisticVQE}, most of recent literature focuses on systems up to 12 qubits \cite{chawla2025trappedionquantumhardware, Shkolnikov2023avoidingsymmetry, Guo2024_ExperimentalQuantumComputationalChemistry}. We hope that our work contributes to increasing the size and variety of molecules studied in the future.

While our methodology has been specifically tailored for VQE, we envision its broader applicability to other fields of material research as a pre-processing step able to make computationally expensive methods, such as Quantum Monte Carlo (QMC), Quantum Phase Estimation (QPE) or Quantum Subspace Expansion (QSE) more feasible as most of them strongly depend on an initial state that is close to the actual ground state. And even beyond molecules or quantum chemistry the classical pre-processing can be useful, for example in Hamiltonian variational methods \cite{Wecker_2015_HamiltonianVariationalMethod}, as it can be applied to any rotation- or excitation based ansätze.

    \section{Methods}    
    The idea of a unitary coupled cluster (UCC) approach reaches back to the eighties and there has been continuous research in the decades since its first mention \cite{Kutzelnigg_1984QuantumChemistryInFockSpace1, Kutzelnigg_1984QuantumChemistryInFockSpace2, Kutzelnigg_1984QuantumChemistryInFockSpace3, Bartlett1988_ExpectationValueCC}. In more recent years, the UCC has been quite popular in the context of quantum computing \cite{Anand_2022QuantumComputingViewOnUCC}, as it is well suited to be mapped to a quantum device due to its unitary nature and because it describes the physically correct states of material systems. The unitary coupled cluster operator can be defined as
    \begin{equation}
        U=e^{-i\sum_N\hat{T}_N},
    \end{equation}
    where $\hat{T}_N=\sum_j\theta_jG_j$ is the sum over generators of fermionic excitations of order $N$, which assume the structure
    \begin{equation}
        {G_{}=i(a_{q_1}^\dagger\ldots a_{q_N}^\dagger a_{p_1}\ldots a_{p_N}-\text{H.c.}}).
    \end{equation}
    The indices $p_1\ldots q_N$ ($q_1\ldots q_N$) denote the $N$ occupied (virtual) orbitals that generator $G$ acts on. Most commonly used is the so-called unitary coupled cluster singles doubles (UCCSD) in which this sum is truncated after the second order ($N=1,2$) of excitation. To map $U$ to a quantum computer however, it must be simplified to not contain a sum in the exponential. This can be achieved by Trotterization \cite{Trotter1959_OnTheProductOfSemi-GroupsOfOperators, Suzuki1976_GeneralizedTrotterFormula}, where $U$ is approximated as
    \begin{equation}
        U\approx\left(\prod_je^{-i\frac{\theta_j}{\tau}G_j}\right)^\tau + \mathcal{O}\left(\frac{1}{\tau}\right),
    \end{equation}
    where an increasing number of Trotter steps $\tau$ leads to better approximations. Trotterization is usually held to one Trotter step~(${\tau=1}$) to keep the required quantum resources to a minimum, which is a good approximation given that the optimal $\theta_j$ are typically small.

    The variational quantum eigensolver (VQE) \cite{Peruzzo2014_VQE} then takes such a unitary
    
    \begin{equation}
        U(\theta)=\prod_jU_j(\theta_j)=\prod_je^{-i\theta_jG_j},
    \end{equation}
    
    and optimizes the tunable parameters $\theta_i$ such that the energy 
    
    \begin{equation}
        E_{\text{VQE}}=\bra{\Psi_0}U^\dagger(\theta) H U(\theta) \ket{\Psi_0}\label{eq:costfunction}
    \end{equation}
    
     is minimized. In ADAPT-VQE \cite{Grimsley2019_ADAPT-VQE}, $U$ is not fixed to contain all excitations from the beginning, but they are iteratively selected from a pool of generators $O=\{G_1, G_2, \dots, G_M\}$ and appended to the ansatz one by one. This comes at the cost of additional measurements for the operator selection, but reduces the depth of the resulting circuit significantly. The energy sorting algorithm (ES) \cite{2023_Fan_energy_sorting} then tries to mitigate this additional overhead by selecting multiple operators at a time.  
    
    \section*{Acknowledgments}
    The authors would like to thank Erik Schultheis for kindly providing code that supported part of the implementation used in this work and for fruitful discussions.
    M.H. acknowledges funding by the DLR Quantum Fellowship Program. 
    This project was made possible by the DLR Quantum Computing Initiative and the Federal Ministry for Economic Affairs and Climate Action; \url{https://qci.dlr.de/quanticom}.

    \section*{Conflicting interests}
    A patent application filed by the German Aerospace Center (Deutsches Zentrum für Luft- und Raumfahrt e.V., DLR), currently pending with the German Patent and Trade Mark Office (Deutsches Patent- und Markenamt, DPMA), covers aspects of this work. It specifically includes, but is not limited to, the combination of the ExcitationSolve and Energy Sorting methods and their application to OVP-CEO operators. The listed inventors are identical to the authors of this work. The application number is DE 10 2025 132 756.4, with the German title \enquote{Verfahren zur Bestimmung von Energien und Energiezuständen eines fermionischen Systems}.
    The authors declare no other financial or non-financial competing interests.
    
    \section*{Data availability}
    The datasets generated and/or analysed during the current study are not publicly available due to their involvement in a pending patent application and associated intellectual property protections. The data can be made available by the corresponding author upon reasonable request.
   
    \section*{Author contributions}
    M.H. led the project, performed all coding and data analysis, and wrote the main draft of the manuscript. T.N.K conducted the theoretical proofs and contributed to the writing and revision of the manuscript. D.B.Y and T.H. contributed to the conceptual development of the work and provided guidance and feedback throughout the research and writing process. All authors reviewed and approved the final manuscript.

    \printbibliography
    
    \onecolumn
\appendix
\title{Efficient Operator Selection and Warm‑Start Strategy for Excitations in Variational Quantum Eigensolvers - Supplementary Material}
\maketitle
\section{Proof that \texorpdfstring{$G^3=G$}{G3=G} holds for OVP-CEOs} \label{sec:proof_CEO_g3=g}

ExcitationSolve requires unitary operators of the form
\begin{equation}
    U(\theta)=\exp{(-i\theta G)},
\end{equation}
with the Hermitian generator $G$ fulfilling $G^3=G$. The two possible OVP-CEO generators are given as 
\begin{align}
    G_{\alpha_1\beta_1\alpha_2\beta_2}^{(\text{OVP-CEO},+)} &\coloneqq G_{\alpha_1\beta_1\rightarrow\alpha_2\beta_2}^{(\textrm{QE})}+G_{\alpha_2\beta_1\rightarrow\alpha_1\beta_2}^{(\textrm{QE})} \nonumber \\
    &=Q_{\alpha_2}^\dagger Q_{\beta_2}^\dagger Q_{\alpha_1}Q_{\beta_1}+Q_{\alpha_1}^\dagger Q_{\beta_2}^\dagger Q_{\alpha_2}Q_{\beta_1}- \textrm{H.c.} \nonumber \\
    &=\frac{1}{4}(X_{\alpha_1}X_{\beta_1}X_{\alpha_2}Y_{\beta_2}-X_{\alpha_1}X_{\beta_1}Y_{\alpha_2}X_{\beta_2}+Y{\alpha_1}Y_{\beta_1}X_{\alpha_2}Y_{\beta_2}-Y_{\alpha_1}Y_{\beta_1}Y_{\alpha_2}X_{\beta_2}) \nonumber \\
    &=\frac{1}{4}X_{\alpha_1}X_{\beta_1}X_{\alpha_2}Y_{\beta_2} \cdot  (I_{\alpha_1}I_{\beta_1}I_{\alpha_2}I_{\beta_2}-I_{\alpha_1}I_{\beta_1}Z_{\alpha_2}Z_{\beta_2}-Z_{\alpha_1}Z_{\beta_1}I_{\alpha_2}I_{\beta_2}+Z_{\alpha_1}Z_{\beta_1}Z_{\alpha_2}Z_{\beta_2}),\\
    G_{\alpha_1\beta_1\alpha_2\beta_2}^{(\text{OVP-CEO},-)} &\coloneqq G_{\alpha_1\beta_1\rightarrow\alpha_2\beta_2}^{(\textrm{QE})}-G_{\alpha_2\beta_1\rightarrow\alpha_1\beta_2}^{(\textrm{QE})} \nonumber \\
    &=Q_{\alpha_2}^\dagger Q_{\beta_2}^\dagger Q_{\alpha_1}Q_{\beta_1}-Q_{\alpha_1}^\dagger Q_{\beta_2}^\dagger Q_{\alpha_2}Q_{\beta_1})- \textrm{H.c.} \nonumber \\
    &=\frac{1}{4}(X_{\alpha_1}Y_{\beta_1}X_{\alpha_2}X_{\beta_2}-X_{\alpha_1}Y_{\beta_1}Y_{\alpha_2}Y_{\beta_2}+Y_{\alpha_1}X_{\beta_1}X_{\alpha_2}X_{\beta_2}-Y_{\alpha_1}X_{\beta_1}Y_{\alpha_2}Y_{\beta_2}) \nonumber \\
    &=\frac{1}{4}X_{\alpha_1}X_{\beta_1}X_{\alpha_2}Y_{\beta_2}\cdot (I_{\alpha_1}Z_{\beta_1}I_{\alpha_2}Z_{\beta_2}-I_{\alpha_1}Z_{\beta_1}Z_{\alpha_2}I_{\beta_2}-Z_{\alpha_1}I_{\beta_1}I_{\alpha_2}Z_{\beta_2}+Z_{\alpha_1}I_{\beta_1}Z_{\alpha_2}I_{\beta_2}),
\end{align}
with the corresponding OVP-CEO unitaries defined as
\begin{equation}
    U_{\alpha_1\beta_1\alpha_2\beta_2}^{(\text{OVP-CEO},\pm)}(\theta)\coloneqq \exp\left(-i\theta G_{\alpha_1\beta_1\alpha_2\beta_2}^{(\text{OVP-CEO},\pm)}\right).
\end{equation}
We now prove that the OVP-CEO generators indeed satisfy the requirement for ExcitationSolve, namely $G^3=G$. We exemplify the proof at hand of $G^+$, but the proof for $G^-$ follows analogously.
We begin with calculating $\left(G^+\right)^2$:
\begin{align}
    \left(G^+\right)^2&=\left[\frac{1}{4}X_{\alpha_1}X_{\beta_1}X_{\alpha_2}Y_{\beta_2} \cdot (I_{\alpha_1}I_{\beta_1}I_{\alpha_2}I_{\beta_2}-I_{\alpha_1}I_{\beta_1}Z_{\alpha_2}Z_{\beta_2}-Z_{\alpha_1}Z_{\beta_1}I_{\alpha_2}I_{\beta_2}+Z_{\alpha_1}Z_{\beta_1}Z_{\alpha_2}Z_{\beta_2})\right]^2 \nonumber \\
    &=\frac{1}{16}(X_{\alpha_1}X_{\beta_1}X_{\alpha_2}Y_{\beta_2})^2 \cdot (I_{\alpha_1}I_{\beta_1}I_{\alpha_2}I_{\beta_2}-I_{\alpha_1}I_{\beta_1}Z_{\alpha_2}Z_{\beta_2}-Z_{\alpha_1}Z_{\beta_1}I_{\alpha_2}I_{\beta_2}+Z_{\alpha_1}Z_{\beta_1}Z_{\alpha_2}Z_{\beta_2})^2 \nonumber \\
    &=\frac{1}{4} (I_{\alpha_1}I_{\beta_1}I_{\alpha_2}I_{\beta_2}-I_{\alpha_1}I_{\beta_1}Z_{\alpha_2}Z_{\beta_2}-Z_{\alpha_1}Z_{\beta_1}I_{\alpha_2}I_{\beta_2}+Z_{\alpha_1}Z_{\beta_1}Z_{\alpha_2}Z_{\beta_2}).
\end{align}
Then, it immediately follows that
\begin{align}
    \left(G^+\right)^3 &= \frac{1}{16} X_{\alpha_1}X_{\beta_1}X_{\alpha_2}Y_{\beta_2}\cdot (I_{\alpha_1}I_{\beta_1}I_{\alpha_2}I_{\beta_2}-I_{\alpha_1}I_{\beta_1}Z_{\alpha_2}Z_{\beta_2}-Z_{\alpha_1}Z_{\beta_1}I_{\alpha_2}I_{\beta_2}+Z_{\alpha_1}Z_{\beta_1}Z_{\alpha_2}Z_{\beta_2})^2 \nonumber \\
    &= \frac{1}{4} X_{\alpha_1}X_{\beta_1}X_{\alpha_2}Y_{\beta_2}\cdot (I_{\alpha_1}I_{\beta_1}I_{\alpha_2}I_{\beta_2}-I_{\alpha_1}I_{\beta_1}Z_{\alpha_2}Z_{\beta_2}-Z_{\alpha_1}Z_{\beta_1}I_{\alpha_2}I_{\beta_2}+Z_{\alpha_1}Z_{\beta_1}Z_{\alpha_2}Z_{\beta_2}) \nonumber \\
    &= G^+,
\end{align}
thus the ExcitationSolve optimization algorithm is applicable.

\clearpage
\section{Simplifying double excitations acting on the Hartree-Fock ground state to Pauli rotations} \label{sec:proof_double_excitation_to_pauli}

To prove that the impact of a double excitation on a classical reference state can efficiently be computed, consider the generator of an arbitrary fermionic double excitation
        \begin{equation}
            G_{pq}^{rs} \coloneqq i(a^\dagger_p a^\dagger_q a_r a_s - \text{H.c}),
        \end{equation} 
        which, under the Jordan-Wigner mapping \cite{jordan1993paulische}, 
        takes the following form
        \begin{align}
            G_{pq}^{rs} \to \frac{1}{8} \mathcal{Z}_{pq}^{rs}  & \left(
            X_pY_qY_rY_s + Y_pX_qY_rY_s - Y_pY_qX_rY_s - Y_pY_qY_rX_s
            \right. \nonumber \\
            & \left. 
            -Y_pX_qX_rX_s - X_pY_qX_rX_s + X_pX_qY_rX_s + X_pX_qX_rY_s
            \right),
        \end{align}
        where the parity string is defined as $\mathcal{Z}_{pq}^{rs}\coloneqq\prod_{j\in\{p, q, r, s\}}\bigotimes_{k<j}Z_k$. Factoring out any of the $XYYY$- or $YXXX$-type odd strings (here we choose $X_p X_q X_r Y_s$), the excitation generators can now be expressed as a product of a Pauli string and some diagonal operator
        \begin{align}
            G_{pq}^{rs} = \frac{1}{8}  \mathcal{Z}_{pq}^{rs}~X_p X_q X_r Y_s & (I_p I_q I_r I_s + I_p I_q Z_r Z_s - I_p Z_q I_r Z_s - I_p Z_q Z_r I_s \\
            &- Z_p I_q I_r Z_s - Z_p I_q Z_r I_s + Z_p Z_q I_r I_s + Z_p Z_q Z_r Z_s).\nonumber
        \end{align}
        This product decomposition is specifically designed such that the $XXXY$ string commutes with the remaining $Z$-terms. This would not be the case if one were to factor out an even string, e.g,~$XXYY$. Now consider the application of the double excitation to the reference state $\ket{\psi}$ where the orbitals $(p,q)$ are occupied and $(r,s)$ are unoccupied:
        \begin{align}
            U_{pq}^{rs}(\theta)\ket{\psi} = \exp(-i\theta G_{pq}^{rs}) \ket{\psi} = \exp(- i \braket{\psi | \mathcal{Z}_{pq}^{rs} | \psi} \theta X_p X_q X_r Y_s) \ket{\psi}.
            \label{eq:double_exc_to_pauli_rot}
        \end{align}
        Exploiting the eigenvalue relation of all $Z$-terms w.r.t.~$\ket{\psi}$ cancels out the pre-factor of $1/8$, and introduces a phase-flip based on the parity of $\ket{\psi}$ w.r.t.~to the orbitals affected by the parity-string $\mathcal{Z}_{pq}^{rs}$.
        The action of the double excitation has therefore been reduced to a single Pauli rotation instead of eight. This can serve as a helpful circuit optimization technique for the first few excitations as it has been demonstrated in Ref.~\cite{Jaeger2025Fast}. 
 
    \section{The energy landscape of double excitations acting on a single Slater determinant} \label{sec:proof_double_excitations_landscape}    
    
        Using the Euler formula, one may now express the resulting variational state as
        \begin{align}
            \ket{\psi(\theta)} = \cos(\theta) \ket{\psi_0} +  \sin(\theta) \braket{\psi_0 | \mathcal{Z}_{pq}^{rs} | \psi_0} \ket{\psi_1},
            \label{eq:double_exc_variational_state}
        \end{align}
        where $\ket{\psi_1}$ differs from $\ket{\psi_0}$ in that the orbitals $(p,q)$ are unoccupied and $(r,s)$ are occupied. Note that this result could be equivalently achieved in the fermionic operator picture without using a fermion-to-qubit mapping, but at the risk of overseeing the circuit simplification we just described. 
        
        In the following supplementary note \ref{sec:proof_classical_simulation_doubles}, we derive equations which describe how much the reference energy can be decreased by the fermionic double excitation. The formula for the energy reduction follows directly from Eq.~\ref{eq:double_exc_variational_state} and reads as follows:
        \begin{align}
            \Delta E(\theta) &= \braket{\psi(\theta)| H | \psi(\theta)} - \braket{\psi_0 | H | \psi_0}  \nonumber 
            \\
            &= \frac{1-\cos(2\theta)}{2} \left[\braket{\psi_1 | H | \psi_1} - \braket{\psi_0 | H | \psi_0}\right] 
            + 
            \frac{\sin(2\theta)}{2} \braket{\psi_0 | \mathcal{Z}_{pq}^{rs} | \psi_0} \left[\braket{\psi_0 | H | \psi_1} + \braket{\psi_1 | H | \psi_0}\right].
        \end{align}
        It is worth pointing out that the derivation presented here for doubles can be generalized towards arbitrary excitations. In particular, for a single excitation $G_p^q$, the parity is $\braket{\psi|\mathcal{Z}_p^q|\psi}$ and the action can be simplified to an $X_pY_q$-type Pauli rotation. For triples and higher orders, the simplification also works. But then the matrix elements $\braket{\psi'|H|\psi}$ coupling the reference state $\ket{\psi}$ to the excited state $\ket{\psi'}$ always vanish and the optimization becomes trivial. In case of a Hartree-Fock calculation, singles, triples and higher will never decrease the initial energy, hence we illustrated the technique solely for doubles. 
        
    \section{Classical simulation of energy sorting for double excitations acting on the Hartree-Fock ground state} \label{sec:proof_classical_simulation_doubles}

        Last, we use the definition of the electronic structure Hamiltonian
        \begin{align}
            H_\mathrm{el.} = \sum_{pq} \sum_\sigma h_{pq}^{\sigma}  a^\dagger_{p^\sigma} a_{q^\sigma} + \frac{1}{2} \sum_{pqrs} \sum_{\sigma \tau} h_{pqrs}^{\sigma \tau}  a^\dagger_{p^\sigma} a^\dagger_{q^\tau} a_{r^\tau} a_{s^\sigma},
        \end{align}
        to compute the matrix elements. Let $\bm o$ denote the set of occupied spin-orbitals in the reference state. Then we can write the reference energy as
        \begin{align*}
            \braket{\psi| H | \psi} 
            = 
            \sum_{p^\sigma \in \bm o} h_{pp}^\sigma 
            +
            \sum_{\{p^\sigma, q^\tau\} \subseteq \bm o} h_{pqqp}^{\sigma \tau} - h_{pqpq}^{\sigma \tau}\delta_{\sigma \tau}.
        \end{align*}
        We define the set of occupied orbitals in the doubly-excited state as $\bm o' = (\bm o \setminus \{i^\mu, j^\nu\}) \cup \{k^\nu, l^\mu\}$, then the same equation with $\bm o'$ holds for the energy of the doubly-excited state.
        We now compute the energy difference between the Hartree-Fock state and the doubly-excited state. In principle one could simply compute these energies separately. However, this involves $\mathcal{O}(n^2)$ two-electron integrals, where $n$ is the number of electrons. Directly evaluating the difference leads to only $\mathcal{O}(n)$ as we illustrate below. In computational practice, this speedup is not particularly relevant, but the structural insights from the result are.
        \begin{align*}
            &\braket{\psi_{i^\mu j^\nu}^{k^\nu l^\mu} | H | \psi_{i^\mu j^\nu}^{k^\nu l^\mu}} - \braket{\psi| H | \psi}  
            \\
            &=
            \sum_{p^\sigma \in \bm o' \setminus \bm o} h_{pp}^\sigma 
            -
            \sum_{p^\sigma \in \bm o \setminus \bm o'} h_{pp}^\sigma 
            +
            \sum_{\{p^\sigma, q^\tau\} \subseteq \bm o' \setminus \bm o} h_{pqqp}^{\sigma \tau} - h_{pqpq}^{\sigma \tau}\delta_{\sigma \tau}
            -
            \sum_{\{p^\sigma, q^\tau\} \subseteq \bm o \setminus \bm o'} h_{pqqp}^{\sigma \tau} - h_{pqpq}^{\sigma \tau}\delta_{\sigma \tau}
            \\
            &=
            h_{kk}^\nu + h_{ll}^\mu + h_{kllk}^{\nu \mu} - h_{klkl}^{\nu\mu}\delta_{\mu\nu} - (h_{ii}^\mu + h_{jj}^\nu + h_{ijji}^{\mu \nu} - h_{ijij}^{\mu \nu} \delta_{\mu\nu})
            \\
            &\hphantom{=~}
            + \sum_{p^\sigma \in \bm o \cap \bm o'} h_{pkkp}^{\sigma \nu} - h_{pkpk}^{\sigma\nu} \delta_{\sigma\nu} + h_{pllp}^{\sigma \mu} - h_{plpl}^{\sigma \mu}\delta_{\sigma\mu} 
            - (h_{piip}^{\sigma \mu} - h_{pipi}^{\sigma \mu}\delta_{\sigma\mu} + h_{pjjp}^{\sigma \nu} - h_{pjjp}^{\sigma \nu} \delta_{\sigma\nu}).
        \end{align*}
        It is worth nothing that the Hartree-Fock ground state is the single slater determinant which minimizes the energy. Hence, any doubly-excited state will have larger energy. The expression from above is therefore strictly positive if $\ket{\psi}=\ket{\psi_\textrm{HF}}$. We then proceed with the Hamiltonian terms which couple the reference state $\ket{\psi}$ to the doubly-excited state $\ket{\psi_{i^\mu j^\nu}^{k^\nu l^\mu}}$. Keep in mind that only 2-body terms can contribute to this coupling. 
        \begin{align}
            &\braket{\psi | \mathcal{Z}_{i^\mu j^\nu}^{k^\nu l^\mu}| \psi} [\braket{\psi | H | \psi_{i^\mu j^\nu}^{k^\nu l^\mu}} + \braket{\psi_{i^\mu j^\nu}^{k^\nu l^\mu} | H | \psi}]
            \nonumber \\
            &= 2 \braket{\psi | \mathcal{Z}_{i^\mu j^\nu}^{k^\nu l^\mu}| \psi}\Re(\braket{\psi_{i^\mu j^\nu}^{k^\nu l^\mu} | H | \psi})
            \nonumber \\
            &= 
            2 \braket{\psi | \mathcal{Z}_{i^\mu j^\nu}^{k^\nu l^\mu}| \psi}
            \Re(\braket{\psi_{i^\mu j^\nu}^{k^\nu l^\mu} | \frac{1}{2}\left\{h_{lkji}^{\mu \nu} + h_{klij}^{\nu \mu} - \left[h_{lkij}^{\mu \nu} + h_{klji}^{\mu \nu}\right]\delta_{\mu\nu} \right\} a^\dagger_{l^\mu} a^\dagger_{k^\nu} a_{j^\nu} a_{i^\mu} | \psi })
            \nonumber \\
            &= 
            \braket{\psi | \mathcal{Z}_{i^\mu j^\nu}^{k^\nu l^\mu}| \psi}
            \braket{\psi_{i^\mu j^\nu}^{k^\nu l^\mu} | a^\dagger_{l^\mu} a^\dagger_{k^\nu} a_{j^\nu} a_{i^\mu} | \psi}
            \Re(2h_{lkji}^{\mu \nu} - \left[h_{lkij}^{\mu \nu} + h_{klji}^{\mu \nu}\right]\delta_{\mu\nu})
            \nonumber \\
            &= \braket{\psi | \mathcal{Z}_{i^\mu j^\nu}^{k^\nu l^\mu}| \psi}^2 
            \Re(2h_{lkji}^{\mu \nu} - \left[h_{lkij}^{\mu \nu} + h_{klji}^{\mu \nu}\right]\delta_{\mu\nu})
            \nonumber \\
            &= \Re(2h_{lkji}^{\mu \nu} - \left[h_{lkij}^{\mu \nu} + h_{klji}^{\mu \nu}\right]\delta_{\mu\nu})
            \nonumber \\
            &= \Re(2h_{ijkl}^{\mu \nu} - \left[h_{jikl}^{\mu \nu} + h_{ijlk}^{\mu \nu}\right]\delta_{\mu\nu}).
            \label{eq:coupling_double_derivation}
        \end{align}
        We note that during an intermediate step the parity reappears and cancels out. This is crucial as the parity is subject to the enumeration of the spin-orbitals (e.g.~different possible realizations of the JW-mapping here), whereas the observables are independent from the mapping. To shorten the notation for the last steps of our calculation, we abbreviate
        \begin{align}
            a_{i^\mu j^\nu}^{k^\nu l^\mu} &= \frac{\braket{\psi_{i^\mu j^\nu}^{k^\nu l^\mu} | H | \psi_{i^\mu j^\nu}^{k^\nu l^\mu}} - \braket{\psi | H | \psi}}{2} \label{eq:a_double}
            \\
            b_{i^\mu j^\nu}^{k^\nu l^\mu} &= \braket{\psi | \mathcal{Z}_{i^\mu j^\nu}^{k^\nu l^\mu}| \psi} \Re(\braket{\psi_{i^\mu j^\nu}^{k^\nu l^\mu} | H | \psi})
            \label{eq:b_double}
        \end{align}
        Then, we rewrite the energy impact as 
        \begin{align}
            \Delta E(\theta) &= a [1-\cos(2\theta)]  + b \sin(2\theta) \nonumber \\
            &= a - \sqrt{a^2+b^2} \cos(2\theta - \arctantwo(-b,a)),
            \label{eq:energy_reduction_landscape}
        \end{align}
        where we dropped the excitation indices for the sake of simplicity.
        With this, we can immediately infer that the largest energy reduction is attained at 
        \begin{align}
            \theta_\textrm{min} = \frac{1}{2}\arctantwo(-b,a),
            \label{eq:optimal_angle}
        \end{align}
        which yields the largest energy reduction at
        \begin{align}
            \Delta E_\textrm{min} \coloneqq \Delta E(\theta_\textrm{min}) = a - \sqrt{a^2+b^2} \leq 0.
            \label{eq:energy_impact}
        \end{align} 
        This way, the impact of any double excitation on single Slater determinant can be efficiently computed.
        Note that for Hartree-Fock calculations, we always have $a>0$ and hence the $\arctantwo(-b, a)$ could be replaced by the regular $\arctan(-b/a)$ function. 
        
        \section{Classical simulation of energy sorting for single excitations acting on a single Slater determinant}  \label{sec:proof_classical_simulation_singles}
        
        If the initial state is not obtained from Hartree-Fock theory (e.g.,~Kohn-Sham theory as observed in UCC-based post-DFT calculations \cite{ma2020quantum, schultheis2025manybody}), single excitations can potentially lower the energy compared to the reference state. In that case the same circuit optimization technique can be applied to reduce the single excitations $U_p^q$ to simple $X_pY_q$-rotations. For the energy difference between the reference state $\ket{\psi}$ and the singly-excited state $\ket{\psi_{i^\mu}^{j^\mu}}$, we find
        \begin{align*}
            \braket{\psi_{i^\mu}^{j^\mu} | H | \psi_{i^\mu}^{j^\mu}} - \braket{\psi| H | \psi}  
            =
            h_{jj}^\mu - h_{ii}^\mu
            + \sum_{p^\sigma \in \bm o \cap \bm o'} h_{pjjp}^{\sigma \mu} - h_{pjpj}^{\sigma\mu}\delta_{\sigma\mu} - (h_{piip}^{\sigma \mu}  - h_{pipi}^{\sigma \mu} \delta_{\sigma \mu}).
        \end{align*}
        Note that unlike the Hartree-Fock ground state, the Kohn-Sham ground state is not guaranteed to be the single Slater determinant which minimizes the energy. Hence the expression can be positive as well. 
        For the coupling between the two states, we can now get contributions from both the one-electron integrals in terms of single excitations, and the two-electron integrals in terms of controlled single excitations:
        \begin{align}
            &\braket{\psi | \mathcal{Z}_{i^\mu}^{j^\mu}| \psi} [\braket{\psi | H | \psi_{i^\mu}^{j^\mu}} + \braket{\psi_{i^\mu}^{j^\mu} | H | \psi}]
            \nonumber \\
            &= 2 \braket{\psi | \mathcal{Z}_{i^\mu}^{j^\mu}| \psi}\Re(\braket{\psi_{i^\mu}^{j^\mu} | H | \psi})
            \nonumber \\
            &= 
            2 \braket{\psi | \mathcal{Z}_{i^\mu}^{j^\mu}| \psi}
            \Re(\braket{\psi_{i^\mu}^{j^\mu} | h_{ji}^{\mu} a^\dagger_{j^\mu} a_{i^\mu} + \frac{1}{2}\sum_{p^\sigma \in \bm o \cap \bm o'} \left\{h_{pjip}^{\sigma \mu} + h_{jppi}^{\mu \sigma} - \left[h_{pjpi}^{\sigma \mu } + h_{jpip}^{\sigma \mu }\right]\delta_{\sigma \mu } \right\} a^\dagger_{p^\sigma} a^\dagger_{j^\mu} a_{i^\mu} a_{p^\sigma}| \psi })
            \nonumber \\
            &= 
            \braket{\psi | \mathcal{Z}_{i^\mu}^{j^\mu}| \psi}
            \braket{\psi_{i^\mu}^{j^\mu} | a^\dagger_{j^\mu} a_{i^\mu} | \psi}
            \Re(2h_{ji}^{\mu} + \sum_{p^\sigma \in \bm o \cap \bm o'} 2h_{pjip}^{\sigma \mu} - \left[h_{pjpi}^{\sigma \mu } + h_{jpip}^{\sigma \mu }\right]\delta_{\sigma \mu })
            \nonumber \\
            &= \braket{\psi | \mathcal{Z}_{i^\mu}^{j^\mu}| \psi}^2 
            \Re(2h_{ji}^{\mu} + \sum_{p^\sigma \in \bm o \cap \bm o'} 2h_{pjip}^{\sigma \mu} - \left[h_{pjpi}^{\sigma \mu } + h_{jpip}^{\sigma \mu }\right]\delta_{\sigma \mu })
            \nonumber \\
            &= \Re(2h_{ji}^{\mu} + \sum_{p^\sigma \in \bm o \cap \bm o'} 2h_{pjip}^{\sigma \mu} - \left[h_{pjpi}^{\sigma \mu } + h_{jpip}^{\sigma \mu }\right]\delta_{\sigma \mu })
            \nonumber \\
            &= \Re(2h_{ij}^{\mu} + \sum_{p^\sigma \in \bm o \cap \bm o'} 2h_{pijp}^{\sigma \mu} - \left[h_{pipj}^{\sigma \mu } + h_{ipjp}^{\sigma \mu }\right]\delta_{\sigma \mu }).
            \label{eq:coupling_single_derivation}
        \end{align}
        The remaining calculations of the optimal angle and energy impact follow the same principles as before. Using the definitions
        \begin{align}
            a_{i^\mu}^{j^\mu} &= \frac{\braket{\psi_{i^\mu}^{j^\mu} | H | \psi_{i^\mu}^{j^\mu}} - \braket{\psi | H | \psi}}{2} 
            \label{eq:a_single}
            \\
            b_{i^\mu}^{j^\mu} &= \braket{\psi | \mathcal{Z}_{i^\mu}^{j^\mu}| \psi} \Re(\braket{\psi_{i^\mu}^{j^\mu} | H | \psi})
            \label{eq:b_single}
        \end{align}
        Then Eq.~\eqref{eq:energy_reduction_landscape} again describes the energy landscape with respect to the parameterized single excitation, and Eqs.~\eqref{eq:optimal_angle}, \eqref{eq:energy_impact} yield the optimal parameter and the largest energy reduction. 

 \newpage   
 
\section{Second order perturbation theory for electronic structure and guess amplitudes} \label{sec:pert_theories}

In this Appendix, we establish the connection between our results from Appendix~\ref{sec:proof_classical_simulation_doubles} and known results from perturbative treatments of electron correlations. We start off with Epstein-Nesbet perturbation theory in second order (EN2) which we find to be closer related to our results, and then also compare with the more widely employed second-order Møller–Plesset perturbation theory (MP2) \cite{Cremer2011}, which has been studied as a selection- and warm-start strategy for UCCSD before \cite{Romero2018Strategies}.

Both EN2 and MP2 are second-order perturbation theories assuming an unperturbed Hamiltonian $H^{(0)}$, which can be exactly solved such that the eigenvalues $E_n^{(0)}$ and eigenvectors $\ket{n^{(0)}}$ are known. In addition, a perturbation $\delta H$ with strength $\lambda$ is considered, such that the goal is to find an approximation to the eigenvalues of eigenvectors of $H=H^{(0)} + \lambda \delta H$. The perturbed energy is then given by
\begin{align}
    E_n(\lambda) = E_n^{(0)} + \lambda \braket{n|\delta H|n} - \lambda^2 \sum_{k\neq n} \frac{|\braket{k|\delta H|n}|^2}{E_k^{(0)}-E_n^{(0)}} + \mathcal{O}(\lambda^3).
    \label{eq:second_order_perturbation}
\end{align}
For the second order correction, we can use that any $\delta H$ (with terms from $H_\textrm{el.}$) can only couple Slater determinants which are related via single- or double excitations. Hence for the perturbative correction to any reference state $\ket{\psi}$ we can write
\begin{align}
    &E_\psi(\lambda) = \braket{\psi|H^{(0)}|\psi} + \lambda \braket{\psi|\delta H|\psi} 
    \nonumber \\
    &- \lambda^2
    \left(
    \sum_{\substack{i^\mu \in \bm o(\psi) \\ j^\mu \in \bm v(\psi)}} \frac{|\braket{\psi_{i^\mu}^{j^\mu}|\delta H|\psi}|^2}{\braket{\psi_{i^\mu}^{j^\mu}|H^{(0)}|\psi_{i^\mu}^{j^\mu}}-\braket{\psi|H^{(0)}|\psi}}
    + 
    \sum_{\substack{\{i^\mu, j^\nu\} \subseteq \bm o(\psi) \\ \{k^\nu l^\mu\} \subseteq \bm v(\psi)}} \frac{|\braket{\psi_{i^\mu j^\nu}^{k^\nu l^\mu}|\delta H|\psi}|^2}{\braket{\psi_{i^\mu j^\nu}^{k^\nu l^\mu}|H^{(0)}|\psi_{i^\mu j^\nu}^{k^\nu l^\mu}}-\braket{\psi|H^{(0)}|\psi}}
    \right) 
    + 
    \mathcal{O}(\lambda^3).
    \label{eq:second_order_perturbation_electronic_structure}
\end{align}
The difference between EN and MP simply boils down to the choice of $H^{(0)}$ and $\delta H$ when treating the electronic structure Hamiltonian. Regardless of the precise choice of $H^{(0)}$ and $\delta H$, we can already take note that all the relevant matrix elements that appear in the perturbative correction have already been calculated in Appendices \ref{sec:proof_classical_simulation_doubles} and \ref{sec:proof_classical_simulation_singles}. To relate the energy from Eq.~\eqref{eq:second_order_perturbation_electronic_structure} to the UCCSD variational parameters, it is then standard practice to define the guess amplitudes (e.g.,~Ref.~\cite{Romero2018Strategies} for doubles with real-valued basis sets)
\begin{align}
    \theta_{i^\mu j^\nu}^{k^\nu l^\mu} 
    &\coloneqq
    -
    \frac{\braket{\psi | \mathcal{Z}_{i^\mu j^\nu}^{k^\nu l^\mu}| \psi}\Re(\braket{\psi_{i^\mu j^\nu}^{k^\nu l^\mu}|\delta H|\psi})}{\braket{\psi_{i^\mu j^\nu}^{k^\nu l^\mu}|H^{(0)}|\psi_{i^\mu j^\nu}^{k^\nu l^\mu}}-\braket{\psi|H^{(0)}|\psi}},
    \label{eq:guess_amplitude_double}
    \\
    \theta_{i^\mu}^{j^\mu} 
    &\coloneqq
    -
    \frac{\braket{\psi | \mathcal{Z}_{i^\mu}^{j^\mu}| \psi}\Re(\braket{\psi_{i^\mu}^{j^\mu}|\delta H|\psi})}{\braket{\psi_{i^\mu}^{j^\mu}|H^{(0)}|\psi_{i^\mu}^{j^\mu}}-\braket{\psi|H^{(0)}|\psi}}.
    \label{eq:guess_amplitude_single}
\end{align}
Note that the real part typically often does not appear explicitly in literature as only real-valued orbitals are considered in many calculations, including the ones in this work.

\subsection{Epstein-Nesbet perturbation theory (EN)} \label{subsec:EpsteinNesbet}
In EN2, one considers the diagonal terms of the electronic structure Hamiltonian $H^{(0)}_\textrm{EN} \coloneqq \textrm{diag}(H_\textrm{el.})$ as the unperturbed Hamiltonian (those terms that entail only number operators), and all the excitation terms (singles, controlled singles, and doubles) as the perturbation $\delta H_\textrm{EN} \coloneqq H_\textrm{el.} - \textrm{diag}(H_\textrm{el.})$. In terms of the electron integrals, we can write
\begin{align}
    H^{(0)}_\textrm{EN} &= \sum_{p} \sum_\sigma h_{pp}^{\sigma}  n_{p^\sigma}
    + 
    \frac{1}{2} \sum_{p\neq q} \sum_{\sigma \tau} (h_{pqqp}^{\sigma \tau}-h_{pqpq}^{\sigma \tau}\delta_{\sigma \tau}) n_{p^\sigma} n_{q^\tau},
    \\
    \delta H_\textrm{EN} 
    &= 
    \sum_{p\neq q} \sum_\sigma h_{pq}^{\sigma} a^\dagger_{p^\sigma} a_{q^\sigma} 
    + 
    \frac{1}{2} \sum_{pqrs} \sum_{\sigma \tau} h_{pqrs}^{\sigma \tau}  a^\dagger_{p^\sigma} a^\dagger_{q^\tau} a_{r^\tau} a_{s^\sigma}
    - 
    \frac{1}{2} \sum_{p\neq q} \sum_{\sigma \tau} (h_{pqqp}^{\sigma \tau}-h_{pqpq}^{\sigma \tau}\delta_{\sigma \tau}) n_{p^\sigma} n_{q^\tau}.
\end{align}
For the zero-th order term, one finds that $\braket{\psi|H^{(0)}_\textrm{EN}|\psi}$ is precisely the reference energy, e.g.,~the Hartree-Fock ground state energy. 
For the first-order term we have $\braket{\psi|\delta H_\textrm{EN}|\psi}=0$ since the perturbation has no diagonal terms. 
For the second-order term, we start with the denominator, where we can simply reuse the definitions for the energy difference between the reference state and the single- and double-excitated states from Eqs.~\eqref{eq:a_double}, and \eqref{eq:a_single}:
\begin{align}
    \braket{\psi_{i^\mu j^\nu}^{k^\nu l^\mu}|H^{(0)}_\textrm{EN}|\psi_{i^\mu j^\nu}^{k^\nu l^\mu}}-\braket{\psi|H^{(0)}_\textrm{EN}|\psi} 
    &= 
    2a_{i^\mu j^\nu}^{k^\nu l^\mu},
    \\
    \braket{\psi_{i^\mu}^{j^\mu}|H^{(0)}_\textrm{EN}|\psi_{i^\mu}^{j^\mu}}-\braket{\psi|H^{(0)}_\textrm{EN}|\psi}
    &=
    2a_{i^\mu}^{j^\mu}.
\end{align}
For the numerator we can readily infer the matrix elements $\braket{\psi'|\delta H_\textrm{EN}|\psi}$ from the definitions in Eqs.~\eqref{eq:b_double} and \eqref{eq:b_single}:
\begin{align}
    \braket{\psi | \mathcal{Z}_{i^\mu j^\nu}^{k^\nu l^\mu}| \psi} \Re(\braket{\psi_{i^\mu j^\nu}^{k^\nu l^\mu} | \delta H_\textrm{EN}  | \psi})
    &= b_{i^\mu j^\nu}^{k^\nu l^\mu}
    \\
    \braket{\psi | \mathcal{Z}_{i^\mu}^{j^\mu}| \psi}\Re(\braket{\psi_{i^\mu}^{j^\mu} | \delta H_\textrm{EN}  | \psi})
    &=
    b_{i^\mu}^{j^\mu}.
\end{align}
Finally, the EN2 guess amplitudes are given by the equations
\begin{align}
    \theta_\textrm{EN2} = -\frac{b}{2a}.
    \label{eq:guess_amplitude_EN}
\end{align}
We can directly relate the EN2 guess amplitudes to the energy sorting amplitudes from Eq.~\eqref{eq:optimal_angle} assuming that the calculation starts from a Hartree-Fock state. Then, as explained earlier, we always have $a>0$, and consequently the EN2 amplitudes arise directly from our exact solution as a first-order Taylor expansion:
\begin{align}
    \theta_\textrm{ES} = \frac{1}{2}\arctantwo(-b, a) \overset{a>0} = -\frac{1}{2}\arctan\left(\frac{b}{a}\right) = -\frac{b}{2a} + \mathcal{O}\left(\left(\frac{b}{a}\right)^2\right)
    = \theta_\textrm{EN2} + \mathcal{O}\left(\left(\frac{b}{a}\right)^2\right).
\end{align}
In a weakly correlated regime, where typically $|b| \ll |a|$, both amplitudes are basically the same. It is only within (strongly) correlated regimes where the amplitudes can significantly differ. In that case, we observe that the Energy Sorting initialization yields better ground state approximations than the EN2 initialization.

\subsection{Møller–Plesset perturbation theory (MP2)} \label{subsec:MøllerPlesset}
In MP2, one considers all (effective) one-electron terms as unperturbed Hamiltonian and the perturbation entails only the genuine two-electrons terms:
\begin{align}
    H^{(0)}_\textrm{MP} &= \sum_{pq} \sum_\sigma \left[h_{pq}^{\sigma} + \sum_{o}\sum_{\tau} \left(h_{pooq}^{\sigma \tau} - h_{poqo}^{\sigma \tau}\delta_{\sigma\tau}\right)n_{o^\tau}\right] a^\dagger_{p^\sigma} a_{q^\sigma} 
    \\
    \delta H_\textrm{MP} &= \frac{1}{2} \sum_{pqrs} \sum_{\sigma \tau} h_{pqrs}^{\sigma \tau}  a^\dagger_{p^\sigma} a^\dagger_{q^\tau} a_{r^\tau} a_{s^\sigma}
    -
    \sum_{pqo} \sum_{\sigma\tau} \left(h_{pooq}^{\sigma \tau} - h_{poqo}^{\sigma \tau}\delta_{\sigma\tau}\right)n_{o^\tau} a^\dagger_{p^\sigma} a_{q^\sigma} 
\end{align}
The unperturbed Hamiltonian $H^{(0)}$ is referred to as an effective one-electron operator because the particle number operators act as scalars with respect to a single Slater determinant, so one can replace $n_{o^\tau}\to 1$ for all occupied spin-orbitals, and $n_{o^\tau}\to 0$ for all unoccupied ones. Then we are left with an one-electron operator. Note however that effective zero-electron terms such as $n_{p^\sigma}n_{q^\tau}$ belong to the perturbation in MP theory. 
For the expectation value of the unperturbed Hamiltonian, one finds
\begin{align}
    E_\textrm{MP}^{(0)} = \braket{\psi|H^{(0)}_\textrm{MP}|\psi} = \sum_{p^\sigma \in \bm o}  h_{pp}^{\sigma} 
\end{align}
which is simply the sum of all occupied orbital energies. Unlike in EN2, the unperturbed Hamiltonian $H^{(0)}_\textrm{MP}$ does not yield the reference energy. However, the reference energy is easily recovered within the first-order correction
\begin{align}
    \braket{\psi|\delta H_\textrm{MP}|\psi} = \sum_{\{p^\sigma, q^\tau\}\subseteq \bm o} h_{pqqp}^{\sigma \tau} - h_{pqpq}^{\sigma\tau}\delta_{\sigma\tau},
\end{align}
such that $E^{(0)}_\textrm{MP} + E^{(1)}_\textrm{MP} = E_\textrm{HF}$ holds for Hartree-Fock calculations. 
Concerning the second order term, we again start with the denominator
\begin{align}
    \braket{\psi_{i^\mu j^\nu}^{k^\nu l^\mu}|H^{(0)}_\textrm{MP}|\psi_{i^\mu j^\nu}^{k^\nu l^\mu}}-\braket{\psi|H^{(0)}_\textrm{MP}|\psi} 
    &= 
    h_{kk}^{\nu} + h_{ll}^{\mu} -  h_{ii}^{\mu} -  h_{jj}^{\mu},
    \\
    \braket{\psi_{i^\mu}^{j^\mu}|H^{(0)}_\textrm{MP}|\psi_{i^\mu}^{j^\mu}}-\braket{\psi|H^{(0)}_\textrm{MP}|\psi}
    &= 
    h_{jj}^{\mu} -  h_{ii}^{\mu},
\end{align}
which corresponds to the difference in orbital energies between the reference and the excited state, but does not capture Coulomb- or exchange interactions anymore unlike in EN2. 
For the matrix elements in the numerator of the second-order term, we obtain
\begin{align}
    \braket{\psi | \mathcal{Z}_{i^\mu j^\nu}^{k^\nu l^\mu}| \psi} \Re(\braket{\psi_{i^\mu j^\nu}^{k^\nu l^\mu} | \delta H_\textrm{MP}  | \psi})
    &= b_{i^\mu j^\nu}^{k^\nu l^\mu}
    \\
    \braket{\psi | \mathcal{Z}_{i^\mu}^{j^\mu}| \psi}\Re(\braket{\psi_{i^\mu}^{j^\mu} | \delta H_\textrm{MP}  | \psi})
    &=
    0.
\end{align}
The matrix element for double excitated states remains that same as in EN2 theory, whereas the coupling between the reference state and single excitated states vanishes in MP theory as the perturbation only entails genuine double excitations. 
Finally, we obtain the well-known MP2 guess amplitudes $\theta_{pq}^{rs}$ given by the equation 
\begin{align}
    (\theta_{i^\mu j^\nu}^{k^\nu l^\mu})_\textrm{MP2} 
    &=
    -
    \frac{b_{i^\mu j^\nu}^{k^\nu l^\mu}}{h_{kk}^{\nu} + h_{ll}^{\mu} -  h_{ii}^{\mu} -  h_{jj}^{\mu}},
    \label{eq:guess_amplitude_double_MP2}
    \\
    (\theta_{i^\mu}^{j^\mu})_\textrm{MP2} 
    &=
    0.
    \label{eq:guess_amplitude_single_MP2}
\end{align}
One key distinction between ES and EN2 is that MP2 cannot be used to predict amplitudes for single excitations. Starting from Hartree-Fock calculations, single excitations cannot reduce the energy anyways, so this is typically not a practical drawback. The other distinction, which we find to be relevant in practice, is that the MP2 guess amplitudes do not incorporate Coulomb energies, which is only a good approximation if $h_{pp}^{\sigma} \gg h_{pqqp}^{\sigma \tau}-  h_{pqpq}^{\sigma\tau}\delta_{\sigma\tau}$. This condition is not necessarily satisfied even in low correlation regimes. 

    \subsection{Numerical comparison}
    
\begin{figure}[hb]
        \centering
        \includegraphics[width=0.9\textwidth]{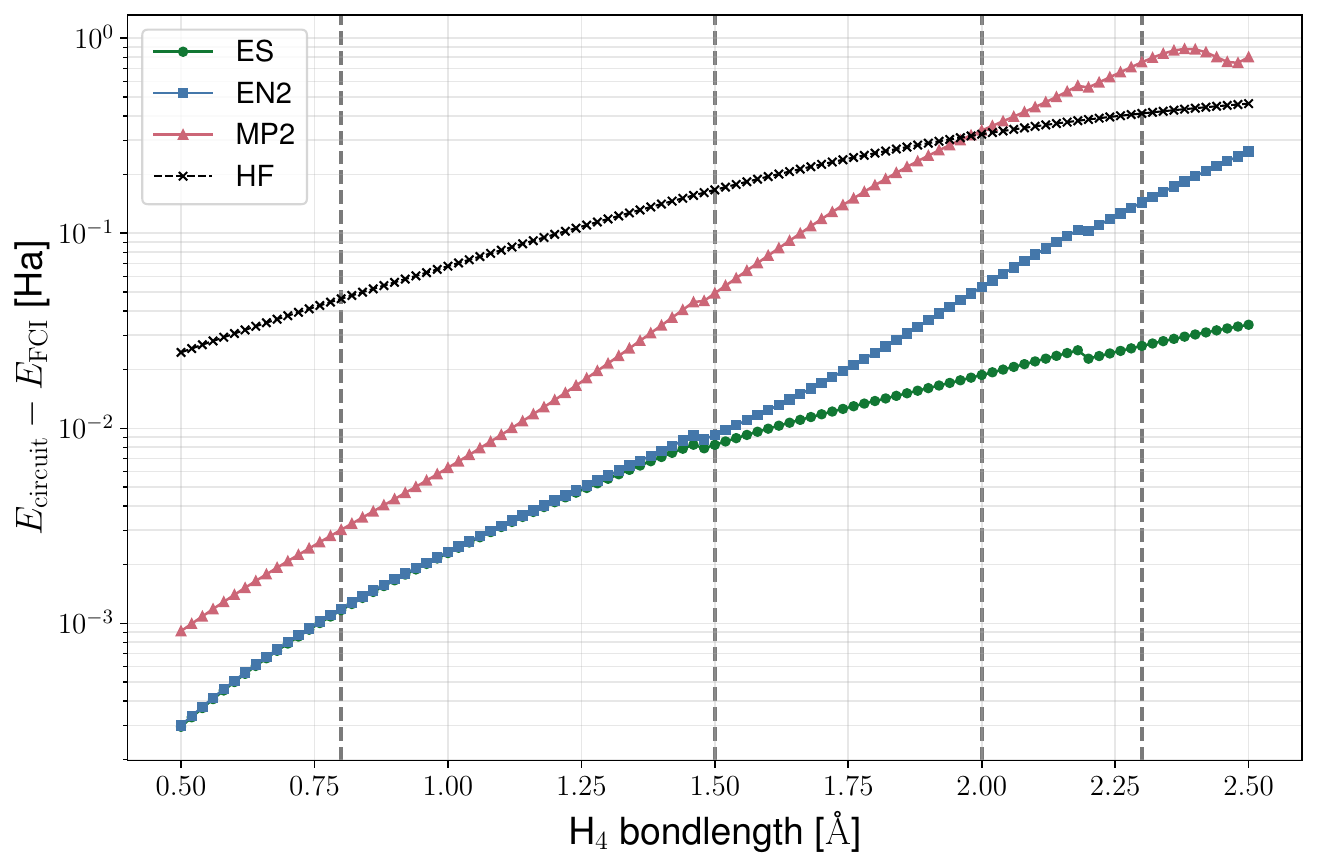}
        \caption{Circuit energies of UCCD circuits across the dissociation curve of \ce{H4} initialized with parameters predicted by MP2, EN2 and ES with HF as reference. Vertical lines mark the bond lenghts for which a full convergence of a UCCDST was performed, shown in \Cref{fig:H4_conv}.}
        \label{fig:H4_diss}
    \end{figure}

\begin{figure}[ht]
        \centering
        \includegraphics[width=0.9\textwidth]{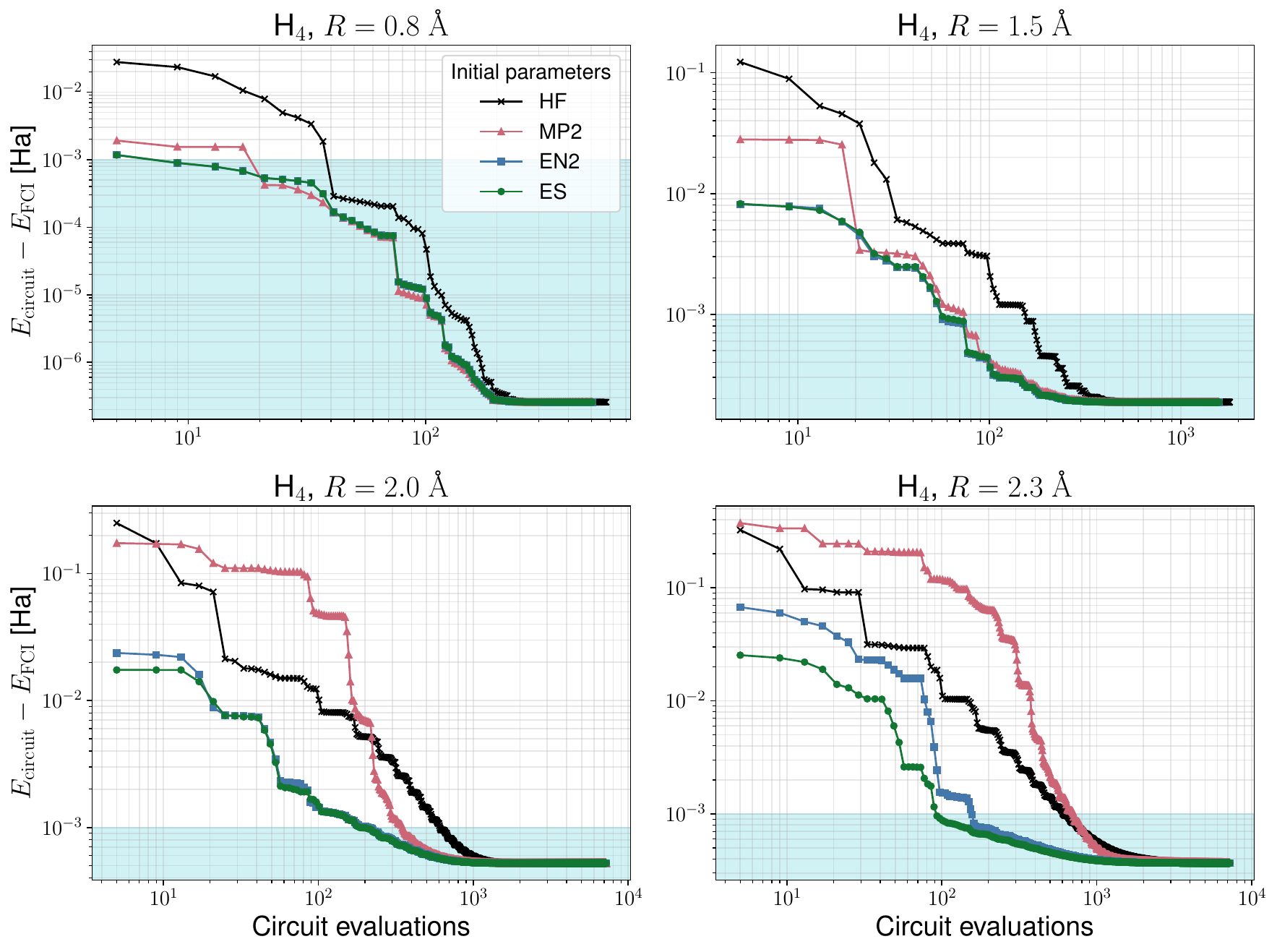}
        \caption{Convergence of UCCSDT for \ce{H4} at different bond distances, where the doubles are initialized with parameters predicted by MP2, EN2 and ES with all-zero initialization (HF) as reference.}
        \label{fig:H4_conv}
    \end{figure}

    Comparing the guess amplitudes $\theta$ predicted by MP2, EN2 and ES shows that EN2 emerges as a first-order Taylor approximation of ES (cf.~Eq.~\eqref{eq:ES_EN2_Amplitudes_Taylor}), which is accurate for weakly correlated systems, but produces a worse initial parameter as the correlation increases, as we demonstrated in \Cref{fig:GuessAmplitudesComparison}. MP2 uses and even stronger approximation and therefore also has a bigger offset even at weak correlations. However a higher energy at initialization does not necessarily mean a slower convergence. We therefore chose four representative bond distances along the \ce{H4} dissociation curve shown in \Cref{fig:H4_diss} where the different optimizers show different behaviors. For each of these bond distances we initialized an UCCD circuit with the optimal parameters chosen by the three methods and with all zeros (HF) as a reference. We then used ES again to select all relevant single- and triple excitation operators and append them to the ansatz. This resulting UCCSDT ansatz was then optimized using ExcitationSolve, the convergence curves are shown in \Cref{fig:H4_conv}. Little surprising, if the parameters are initialized at similar values, the convergence follows almost the same curve. But even when deviations are large as is the case for MP2 compared to EN2 and ES at \SI{1.5}{\angstrom}, the re-evaluation of a few parameters suffices to correct the error. Only initializing at all zeros consistently shows slower convergence, even when the MP2 initialization starts out at a higher energy. So while ES provides a more reliable initial guess of the parameters even at strong correlations, in the end it saves little in computational costs when running a full VQE optimization to convergence. Only in the very extreme case of completely dissociated atoms can the initialization show a small advantage in reaching chemical accuracy.

\section{Operator selection for other bondlengths} \label{sec:op_selection_bondlengths}
    
\begin{figure}[hb]
        \centering
        \includegraphics[width=0.9\textwidth]{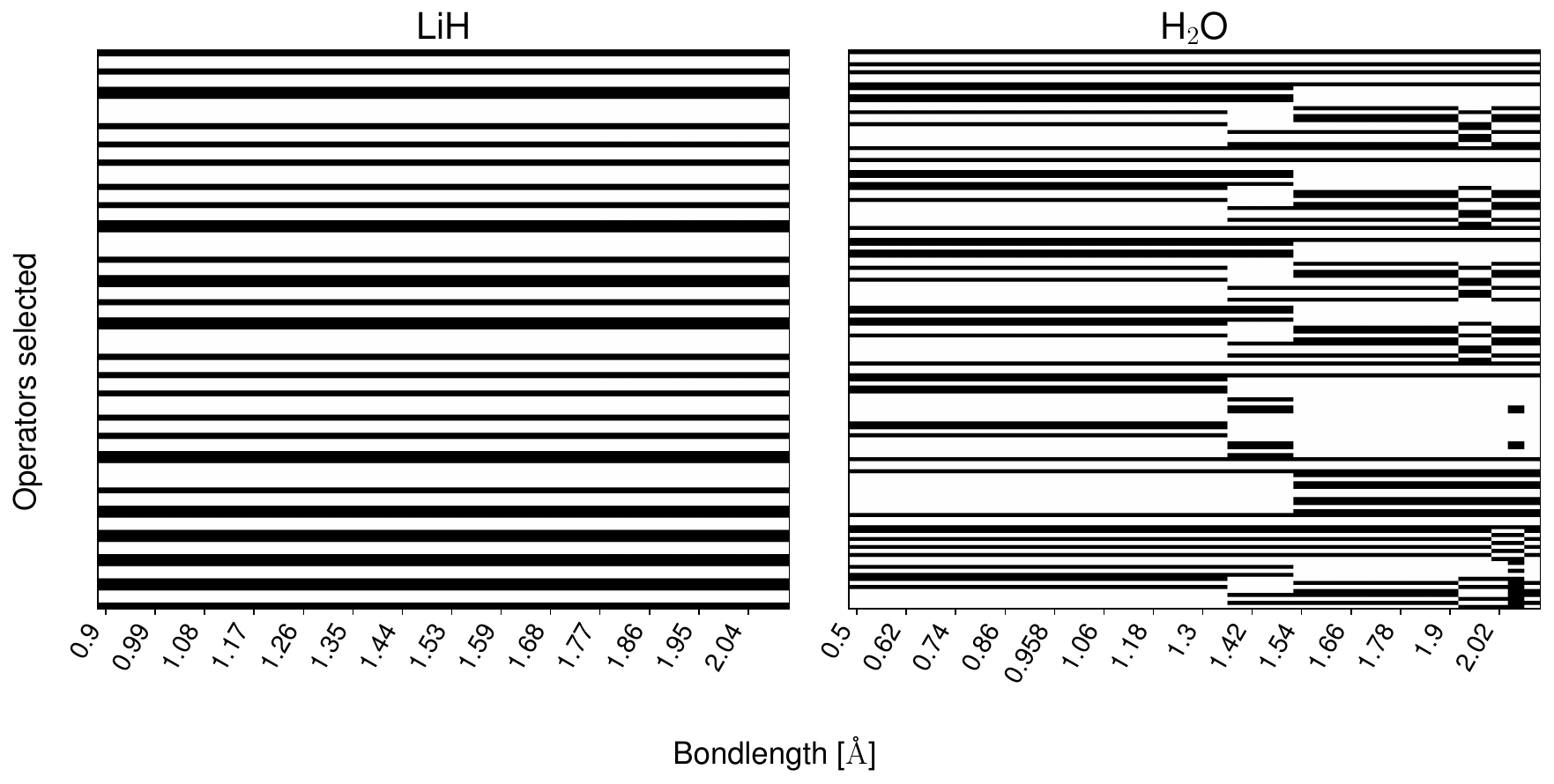}
        \caption{Operator selection for \ce{LiH} and \ce{H2O} using ExcitationSolve and ES. Operators marked black are selected for that bondlength, operators marked in white are not relevant. The y-axis shows all the operators in the operator pool.}
        \label{fig:op_selection}
    \end{figure}

    We find that for molecules in a linear alignment such as the \ce{LiH} molecule the same operators are selected by ES for all bondlengths. In \Cref{fig:op_selection} we plot, which operator is selected at different bondlengths for the \ce{LiH} and \ce{H2O} molecules, black meaning that the operator is chosen, white that it is not. Without going into detail which operators are the relevant ones, it becomes apparent, that for \ce{LiH} the bondlength does not play a role for the selection. For \ce{H2O} on the other hand the selected operators vary with the bondlength. The reason is that in linear molecules even as the bond is stretched or compressed, the same orbitals are overlapping and so the same excitations are relevant. In a molecule such as \ce{H2O} on the other hand with a bond angle of $104.5\degree$ the overlap between orbitals changes and so also the operator selection is dependent on the bondlength. However even in \ce{H2O} many similar bondlengths select similar operators. Therefore if we want to know the relevant operators at multiple bondlenghts it may only be necessary to determine them once (for linear molecules) or a few times (for non-linear molecules) and interpolate the rest saving further computational power.
    
\end{document}